\pdfoutput=1
\documentclass[english,11pt,onecolumn]{IEEEtran}
\usepackage[T1]{fontenc}
\usepackage{babel}

\usepackage{float}
\usepackage{amsthm}
\usepackage{amsmath}
\usepackage{graphicx}
\usepackage{amssymb}

\makeatletter

\theoremstyle{plain}
\newtheorem{thm}{Theorem}
\theoremstyle{plain}
\newtheorem{cor}[thm]{Corollary}

\makeatother

\begin{document}

\title{A Frame Rate Optimization Framework For Improving Continuity In Video
Streaming}

\author{Evan Tan%
\thanks{E. Tan is with the School of Computer Science and Engineering, University
of New South Wales, Sydney NSW 2052, Australia (email: evant@cse.unsw.edu.au)%
} and Chun Tung Chou%
\thanks{C. T. Chou is with the School of Computer Science and Engineering,
University of New South Wales, Sydney NSW 2052, Australia (email:
ctchou@cse.unsw.edu.au)%
}}
\maketitle
\begin{abstract}
This paper aims to reduce the prebuffering requirements, while maintaining
continuity, for video streaming. Current approaches do this by making
use of adaptive media playout (AMP) to reduce the playout rate. However,
this introduces playout distortion to the viewers and increases the
viewing latency. We approach this by proposing a frame rate optimization
framework that adjusts both the encoder frame generation rate and
the decoder playout frame rate. Firstly, we model this problem as
the joint adjustment of the encoder frame generation interval and
the decoder playout frame interval. This model is used with a discontinuity
penalty virtual buffer to track the accumulated difference between
the receiving frame interval and the playout frame interval. We then
apply Lyapunov optimization to the model to systematically derive
a pair of decoupled optimization policies. We show that the occupancy
of the discontinuity penalty virtual buffer is correlated to the video
discontinuity and that this framework produces a very low playout
distortion in addition to a significant reduction in the prebuffering
requirements compared to existing approaches. Secondly, we introduced
a delay constraint into the framework by using a delay accumulator
virtual buffer. Simulation results show that the the delay constrained
framework provides a superior tradeoff between the video quality and
the delay introduced compared to the existing approach. Finally, we
analyzed the impact of delayed feedback between the receiver and the
sender on the optimization policies. We show that the delayed feedbacks
have a minimal impact on the optimization policies.
\end{abstract}

\section{Introduction}

Video continuity is the length of time the video is being played without
interruptions. A low video continuity would result in a stop-start
video that is known to impact on the viewer perceived video quality
\cite{ref:jitterperceptual}. One of the main causes of video discontinuity
in streaming video is due to buffer underflow caused by for example
network delay. The traditional way to reduce the buffer underflow
occurrences is to prebuffer video at the decoder. The prebuffering
amount needs to be sufficiently large to maintain video continuity.
However, a large prebuffering introduces unwanted initial delays into
the system. 

To reduce the amount of prebuffering required while maintaining video
continuity, current approaches make use of adaptive media playout
(AMP) \cite{ref:adaptiveamp1,ref:ampkalman,ref:ampsmoother3}. AMP
approaches reduce the \emph{playout frame rate} of the decoder in
order to avoid buffer underflows, this is because slowing down the
playout frame rate is preferable to halting the playout \cite{ref:scaleisbetter}.
The playout frame rate is defined as the rate of frames being removed
from the decoder buffer for decoding and playout to the viewer. AMP
has been shown to reduce prebuffering while maintaining video continuity
\cite{ref:ampkalman}. However, it introduces \emph{playout distortion}
to the viewers, this is because the video is being played slower than
its natural playout frame rate (also known as the video capture frame
rate). Furthermore, AMP schemes adjust the playout frame rate independent
of the encoder strategy, this potentially introduces more playout
distortion than required. While the playout distortion can be reduced
by limiting the playout slowdown, it potentially affects the video
continuity. Moreover, slowing down the playout frame rate increases
the viewing latency and introduces additional delay into streaming
system.

In this paper, we aim to reduce the amount of prebuffering required
by dealing with the continuity of the video in a reactive manner.
We realize this aim by proposing a framework that performs frame rate
control at both the encoder and decoder as well as introduce a way
to constrain the viewing latency. The key idea is that, if the network
bandwidth drops and the number of frames in the decoder buffer is
low, the encoder should send more frames to the decoder to prevent
buffer underflow from occurring, thus maintaining playout continuity.
In order to approach this systematically, we formulate an optimization
problem that takes into account the video continuity, video quality
and overall playout delay. The contributions of this paper are:
\begin{enumerate}
\item We propose a frame rate control framework that jointly adjusts the
encoder frame generation rate and the playout frame rate. This distinguishes
our framework from conventional approaches that do not perform frame
rate control and AMP approaches that only adjusts the playout frame
rate.
\item We derive an optimization formulation of the frame rate control framework
using the technique of virtual buffer. We then use Lyapunov optimization
\cite{ref:neelybook} on this model to systematically derive the optimization
policies. We show that these policies can be decoupled into separate
encoder and decoder optimization policies with feedback between the
two. This allows for a distributed implementation of the policies.
We demonstrate that this framework produces a very low playout distortion
in addition to a significant reduction in the prebuffering requirements
compared to existing approaches.
\item A delay constraint that reduces the accumulated delay from playout
slowdowns. We then show that the delay constrained framework provides
a superior tradeoff between the video quality and the delay introduced
compared to the existing approach.
\item An analysis of the impact of delayed feedback between the receiver
and the sender. We show that the delayed feedbacks have a minimal
impact on the optimization policies. 
\end{enumerate}
This paper is organized as follows. Section \ref{sec:Related-Work}
reviews the current approaches. Section \ref{sec:Discontinuity-Penalty}
demonstrates how the frame rate control problem can be modelled. Section
\ref{sec:Delay-Constrained-Frame} demonstrates how a delay constraint
can be introduced into the framework. Section \ref{sec:Network-Delay-Impact}
analyzes the delayed feedback on the optimization policies. Section
\ref{sec:Video-Quality-Functions} describes the video quality functions
that can be used with this framework. Section \ref{sec:Performance-Evaluation}
evaluates the performance of the proposed improved framework. Section
\ref{sec:Conclusions} concludes this paper.

\section{Related Work\label{sec:Related-Work}}

Prebuffering video data has been studied in \cite{ref:bufcalc2,ref:bufcalc1,ref:bufcalc3}.
These techniques are focused on calculating the correct amount of
prebuffered data based on a mathematical model to avoid the occurrence
of a buffer underflow once the video playout has started. However,
these techniques do not take into account of the varying playout rates
and the resulting reduction in prebuffering into their models. Furthermore,
prebuffering introduces unwanted delay into the system and is known
to have an impact on the user perceived quality of the video \cite{ref:delaynjitterimpact}.

Sender rate adaptation techniques such as encoder rate control \cite{ref:h264rc,ref:x264rc}
and scalable rate control\cite{ref:h264scalable} achieve video continuity
by ensuring the video rate matches the network bandwidth. However,
these approaches do not take into account of the network delay variation
that might occur (e.g. due to network jitter). Moreover, it has been
demonstrated that coupling AMP with sender rate adaptation techniques
can further reduce the prebuffering requirements \cite{ref:ampkalman}. 

AMP is another class of techniques that is used to improve video continuity.
The main idea behind AMP is to reduce the playout rate of the received
media. This reduces the departure rate of the media in the decoder
buffer and potentially allows the buffer to be filled up to safer
levels. 

AMP has been studied extensively for audio applications, \cite{ref:audioamp1,ref:audioamp2,ref:audioamp3,ref:audioamp4,ref:audioamp5,ref:audioamp6}.
These audio AMP techniques depend on audio scaling techniques such
as WSOLA \cite{ref:wsola}, which allows audio to be scaled without
changing its pitch. A recent work by Damnjanovic et al has shown that
audio and video scaling can be done in real-time \cite{ref:audioscale}.
In this paper, we focus on AMP for video.

AMP for video involves scaling the frame intervals to slowdown or
speedup the playout rate. The slowdown or speedup is triggered by
a threshold set on the buffer occupancy. Once the buffer occupancy
drops below the threshold, AMP will slow down the playout rate and
vice versa. There have been studies conducted on the dynamic adjustment
of this threshold \cite{ref:adaptiveamp1,ref:adaptiveamp2,ref:adaptiveamp3,ref:ampseo}.
The adjustment is normally based on the network condition, the poorer
the network condition, the higher the threshold. These techniques
mainly based the threshold on the buffer occupancy. We instead record
the difference between the receiving frame interval and the playout
frame interval into a virtual buffer and treat it as a penalty, which
is equivalent to soft thresholding.

AMP has also been integrated into the design of packet schedulers
\cite{ref:ampscheduler1,ref:ampscheduler2,ref:ampscheduler3,ref:ampscheduler4,ref:ampscheduler5}.
These techniques tend to slowdown the playout rate for important video
packets. This ensures that the more important video packets have a
higher probability of meeting its deadline and thus avoid being dropped.
We do not focus on packet scheduling in this paper and our proposed
framework could complement any packet scheduling scheme.

Another aspect of AMP being studied is the smoothness of transition
between playout rate adjustment \cite{ref:ampsmoother3,ref:ampsmoother1,ref:ampsmoother2}.
The goal of these approaches is to ensure that adjustments made to
the playout rate is done as smoothly as possible so as to reduce any
noticeable effects to the viewers. We do not focus on rate smoothing
in the current paper and again any smoothing scheme can be used within
the framework.

Steinbach et al \cite{ref:ampsteinbach} and Kalman et al \cite{ref:ampkalman}
both examined the trade-off between delay and buffer underflow using
a two state Markov models. Kalman et al further proposed AMP-Initial,
AMP-Robust and AMP-Live. AMP-Initial slows down the playout rate until
the buffer reaches a certain target level, this produces a lower perceived
initial buffering delay to the viewer. AMP-Robust slowdowns the playout
rate if the current buffer occupancy falls below the target buffer
level while AMP-Live slowdowns or speedups the playout rate to maintain
the target buffer level.

These AMP approaches mainly examine only the effects of adjusting
the playout frame rate independently and do not consider any encoder
strategy to reduce playout distortion. While our proposed approach
aims to examine the effects on video quality and viewing delay based
on the adjustment of both the encoder frame generation rate and the
playout frame rate. Slowing down the playout frame rate also introduces
viewing latency, we will propose a way to constrain this latency in
our proposed approach.

\section{Frame Rate Control For Video Continuity}

\begin{figure}[H]
\includegraphics[scale=0.43]{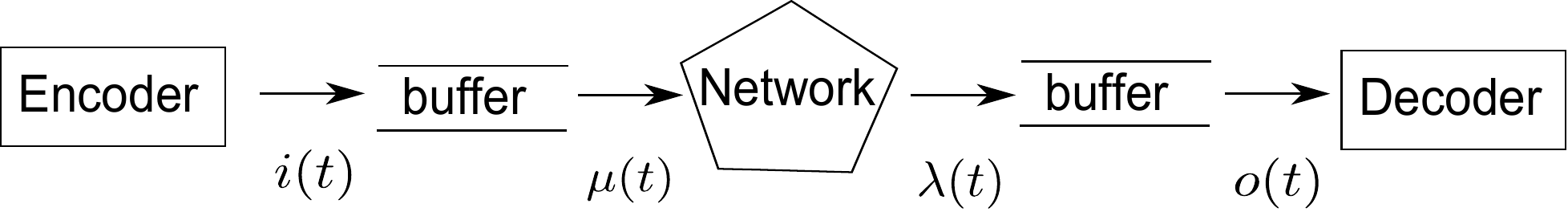}\caption{Encoding-decoding flow with encoder frame generation rate $i(t)$,
sending frame rate $\mu(t)$, receiving frame rate $\lambda(t)$ and
playout frame rate $o(t)$. All rates are in frames per seconds.\label{fig:encdec}}

\end{figure}

Figure \ref{fig:encdec} shows a typical video transmission scenario.
The encoder generates video frames at a rate of $i(t)$ frames per
second (fps) at time $t$ into the encoder buffer. The network transport
protocol then transmits the video data from the encoder buffer into
the network at a rate of $\mu(t)$ fps. The transmitted video data
will be received at the decoder at a rate of $\lambda(t)$ fps in
its buffer. The decoder then proceeds to playout the received video
data from the decoder buffer at a rate of $o(t)$ fps. Video data
arriving after the playout deadline are assumed to be lost. In the
scenario described in fig. \ref{fig:encdec}, there are two mechanisms
that can be used to maintain video continuity: reducing playout frame
rate $o(t)$ and increasing the encoder frame generation rate $i(t)$. 

Our analytical model takes network delay into consideration but does
not consider packet losses %
\footnote{Note that even though we do not consider packet loss explicitly, the
simulation results in section \ref{sec:Performance-Evaluation} show
that our optimisation framework works well in the presence of packet
loss.%
}. In particular, an important factor that affects the performance
is the inter-frame delay, which is defined as the difference of network
delays for consecutive video frames. We assume that the frames are
received into the decoder buffer in display order, so reordering is
done before the frames enter the decoder buffer. Thus, inter-frame
delays can be affected by network jitter and packet losses. Notice
that this model allows us to focus on calculating the amount of frames
arriving into and departing from the decoder buffer such that decoder
buffer underflow can be avoided, and video continuity can be maintained.
We also assume, in this paper, that the encoder is able to generate
frames faster than the natural playout frame rate. This is a reasonable
assumption since there are existing encoders such as x264 \cite{ref:x264}
that can encode frames in real-time faster than a typical natural
playout frame rate of 30 frames per second (fps).

Slowing down the video playout frame rate $o(t)$ is termed in the
literature as AMP. The idea is that the slower video playout allows
time for the buffer to fill up to the required level without the need
to stop the video for rebuffering. 

Another way to maintain video continuity is to increase the \emph{encoder
frame generation rate} $i(t)$, which refers to the amount of frames
the encoder actually sends out into the network and it is important
to point out that the encoder frame generation rate is \emph{not}
a temporally scaled frame rate. Note that we purposely chose the term
encoder frame generation rate to differentiate it from the commonly
used term of encoder frame rate because they represent two \emph{different}
concepts. The meaning of encoder frame generation rate is best illustrated
by an example: when no frame rate control is used, the encoder frame
generation rate $i(t)$ is the natural playout frame rate. Let us
assume that to be 30 fps; when frame rate control is used, the encoder
may increase the encoder frame generation rate $i(t)$ to say 60 fps
to quickly fill up the decoder buffer to maintain continuity. Note
that these 60 frames are still encoded using the same natural playout
frame rate of 30 fps, so they form the next 2 seconds of video. This
example illustrates that a higher encoder frame generation rate means
more than one second of video is generated in one second but the video
is always encoded using the same natural playout frame rate. 

Increasing $i(t)$ allows potentially more frames to reach the decoder
and increase the decoder buffer level. This, in turn, helps to improve
the continuity of the video. However, increasing $i(t)$ is likely
to cause the video bitrate to increase as more frames are produced
per second by the encoder. To ensure that the video bitrate does not
exceed the available bandwidth, we introduce additional compression
to the video such that the higher the encoder frame generation rate
$i(t)$, the higher the compression applied. 

For a given encoder generation rate $i(t)$, the higher compression
is obtained by the rate controller adjusting the encoder such that
the average frame size produced by the encoder is:

\begin{equation}
\textrm{average frame size}=r(i(t))=\frac{ABR(t)}{i(t)}\label{eq:fropt_avgfrmsiz}\end{equation}

where $ABR(t)$ is the available bandwidth%
\footnote{ideally the encoder should use $ABR(t+d_{s})$, where $d_{s}$ is
the sender buffer delay. However, since this is difficult to determine,
we use $ABR(t)$ to approximate $ABR(t+d_{s})$.%
} at time $t$. In practice, the average frame size produced by a rate
control strategy may not satisfy \eqref{eq:fropt_avgfrmsiz}. However,
\eqref{eq:fropt_avgfrmsiz} is what most rate control schemes try
to meet as they use \eqref{eq:fropt_avgfrmsiz} as part of their fluid
flow model to determine the amount of bits to allocate to a frame
\cite{ref:h264rc,ref:x264rc,ref:mpeg4rc}. 

It can be shown that when \eqref{eq:fropt_avgfrmsiz} is satisfied
by the rate controller, the resulting video bitrate will not exceed
the available bandwidth for different values of the encoder frame
generation rate $i(t)$. Also, the average frame size decreases when
$i(t)$ is increased which would tend to reduce the frame quality.
Finally, a higher $i(t)$ also allows more frames to be sent to the
decoder, thus it allows us to increase the buffer level and maintain
video continuity.

Reducing $o(t)$ and increasing $i(t)$ increases playout distortion
and reduces frame quality respectively while improving video continuity.
This suggests that an optimal trade-off need to be found. To do this,
we model the frame rate control problem as an optimization problem
and make use of Lyapunov optimization to obtain policies that help
determine the optimal trade-off.

In this paper, we adjust the encoder frame generation rate $i(t)$
and playout frame rate $o(t)$ by adjusting the encoder frame generation
interval $\frac{1}{i(t)}$ and the playout frame interval $\frac{1}{o(t)}$
respectively. The reason we chose to adjust the intervals is because
it allows the optimization problem to be concave and, therefore, easier
to solve. This issue will be discussed in more detail in Section \ref{sec:Video-Quality-Functions}.

\section{Discontinuity Penalty For Frame Rate Optimization\label{sec:Discontinuity-Penalty}}

\subsection{Buffering Criteria}

\begin{figure}[H]
\begin{centering}
\includegraphics[scale=1.1]{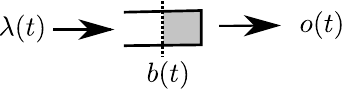}
\par\end{centering}

\caption{Receiver buffer model.}
\label{fig:frmmodel}
\end{figure}

Since the video discontinuity is correlated to decoder buffer underflows,
we first study how buffer underflow occurs. To do this, we make use
of the receiver model illustrated in fig. \ref{fig:frmmodel}. We
assume that the sender, network and receiver all work in slotted time.
At time slot $t$, the receiver receives $\lambda(t)$ frames (receiving
frame rate) and stores it into the receiver buffer. Simultaneously,
$o(t)$ frames are being removed from the buffer and played out to
the viewer (playout rate) at time $t$. Let $b(t)$ be the amount
of buffered video frames in the buffer at time $t$ and $T$ be the
length of the sequence in time slots, then to avoid an underflow the
following condition needs to be met: 

\begin{equation}
b(t)\geq\sum_{\tau=t}^{T}o(\tau)-\sum_{\tau=t}^{T}\lambda(\tau)\quad\quad\forall T\geq t\label{eqn:b0}\end{equation}

Equation \eqref{eqn:b0} intuitively means that, to avoid a buffer
underflow, the cumulative buffer drainage for the rest of the sequence
playing time should not exceed the current buffer occupancy $b(t)$. 

Now we refine the above model to make use of \emph{frame intervals}.
This is done to build up a system model using frame intervals. Basically
to make use of the frame intervals, we set $\lambda(t)=\frac{1}{r(t)}$
and $o(t)=\frac{1}{p(t)}$. Equation \eqref{eqn:b0} then becomes:

\begin{align}
b(t) & \geq\sum_{\tau=t}^{T}\frac{r(\tau)-p(\tau)}{r(\tau)\, p(\tau)}\label{eqn:b1}\end{align}

We assume that $r_{min}\leq r(t)\leq r_{max}$ and $p_{min}\leq p(t)\leq p_{max}$,
i.e. both $r(t)$ and $p(t)$ are bounded. That would mean that we
can approximate \eqref{eqn:b1} as:

\begin{align}
b(t)\, r_{min}\, p_{min} & \geq\sum_{\tau=t}^{T}r(\tau)-p(\tau)\label{eqn:b2}\end{align}

Note that the choice of using $r_{min}$ and $p_{min}$ results in
a more conservative bound. An alternative bound, which is looser,
is to replace the left-hand-side of \eqref{eqn:b2} by $b(t)\, r_{max}\, p_{max}$.
We will show in simulation results (section \ref{sec:Performance-Evaluation})
that these two choices give similar results.

If we divide \eqref{eqn:b2} by $T-t$, the remaining time slots left,
we can estimate the buffer underflow bound \emph{per time slot}. This
will result in: 

\begin{equation}
\frac{b(t)\, r_{min}\, p_{min}}{T-t}\geq\hat{r}-\hat{p}\label{eqn:b3}\end{equation}

where $\hat{r}$ and $\hat{p}$ are the averages of $r(t)$ and $p(t)$
respectively. Equation \eqref{eqn:b3} provides us with a way to design
the optimization policy to avoid buffer underflow in a time slot.
Since Lyapunov optimization works on a per time slot basis, we prefer
\eqref{eqn:b3} over \eqref{eqn:b2}. Essentially, we need to design
a policy that produces a \emph{receiving frame interval} $r(t)$ and
a \emph{playout frame interval} $p(t)$ in such a way that the above
bound is met.

\subsection{System Model\label{sub:fro-imp-System-Model}}

\begin{figure}[H]
\includegraphics[scale=0.5]{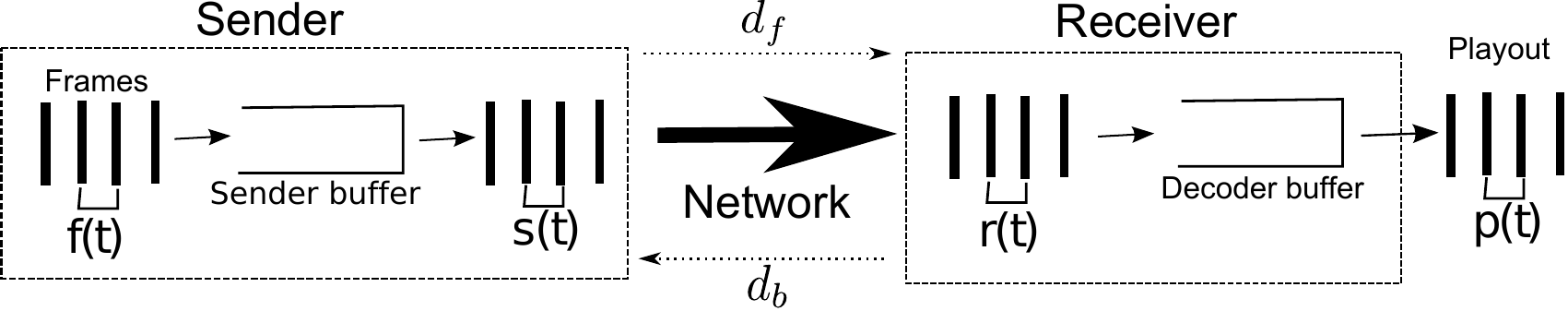}

\caption{System model showing the frame intervals.}
\label{fig:sysmodel}
\end{figure}

We now show how $r(t)$ and $p(t)$ are produced in the complete system
model. Fig. \ref{fig:sysmodel} illustrates the system model. In a
time slot $t$, the encoder at the sender produces video frames at
intervals of $f(t)$ and stores them into the sender buffer. The sender
sends the video frames in its buffer at intervals of $s(t)$. Note
that $f(t)$ and $s(t)$ are the \emph{encoder frame generation interval}
and the \emph{sending frame interval} respectively. Simultaneously,
the receiver receives video frames from the sender at intervals of
$r(t)$ and puts them into the decoder buffer. The decoder in the
receiver plays out the video frames at intervals of $p(t)$ to the
viewer, $p(t)$ is the \emph{playout frame interval}. The network
will also produce a forward delay of $d_{f}$ and a backward delay
of $d_{b}$.

The goal of our framework is to jointly adjust the encoder frame generation
interval $f(t)$ and the playout frame interval $p(t)$ to maintain
video continuity. To simplify the model, we assume that $f(t)=s(t)$,
while this essentially assumes no delay caused by the sender buffer,
delays caused by the sender buffer is simulated in our experiments
later on. We also assume that $f(t)$ is bounded within the range
$[f_{min},f_{max}]$ and $r(t)$ is defined by a network delay variation
function $F(s(t))$ as : $r(t)=F(s(t))=F(f(t))$, since $f(t)=s(t)$. 

This means that we can represent the delay variation based on the
encoder frame generation interval $f(t)$. In this paper, we specify
$F(f(t))$ as:

\begin{equation}
F(f(t))=e(t)\times f(t)\label{eqn:Fdefine}\end{equation}

Where $e(t)$ is the frame interval scaling factor due to delay variations
from the network. In practice, we estimate $e(t)$ at the receiver
by:

\begin{equation}
e(t)=\frac{r(t)}{f(t-d_{f})}\label{eqn:delayscale}\end{equation}

Where $d_{f}$ is the forward delay between the sender and receiver.
Note that if there is no delay (i.e. $d_{b}=d_{f}=0$), it will mean
that $F(f(t))=r(t)$.

\subsection{General Optimization Problem\label{sub:General-Optimization-Problem}}

There are three main objectives that we want to optimize: 1. frame
quality, 2. playout distortion and 3. continuity. Frame quality is
defined as the perceived visual quality of the video. This will be
represented as a frame quality function $g(f(t))$, where $g(f(t))$
is an increasing function of $f(t)$. Playout distortion is the perceived
distortion when the playout rate deviates from the natural frame rate
and will be represented by a function $h(p(t))$, where $h(p(t))$
is a convex function of $p(t)$. Both $g(f(t))$ and $h(p(t))$ are
non-negative functions, i.e. $g(f(t))\geq0$ and $h(p(t))\geq0$,
and are assumed to be uncorrelated. We will suggest a specific form
for $g(f(t))$ and $h(p(t))$ later in Section \ref{sec:Video-Quality-Functions}.
Continuity is the length of time the video is played without interruptions
due to buffer underflow. We ensure continuity in this framework by
ensuring that a \emph{virtual buffer stabilizes}, this concept will
be explained further in the next section.

With the system model described in the previous section, we now formulate
a general optimization problem:

\begin{align}
\mathrm{\textrm{Maximize:}} & \quad g(f(t))-h(p(t))\label{eqn:lyapoptobj}\\
\textrm{Subject to:} & \quad U(t)\textrm{ is stable}\label{eqn:lyapoptobj1}\\
 & \quad f_{min}\leq f(t)\leq f_{max}\label{eqn:lyapoptobj2}\\
 & \quad p_{min}\leq p(t)\leq p_{max}\label{eqn:lyapoptobj3}\end{align}

Where \emph{$U(t)$} is the virtual buffer representing the \emph{discontinuity
penalty}. Since the objective \eqref{eqn:lyapoptobj} is separable,
maximizing \eqref{eqn:lyapoptobj} can be seen as maximizing the frame
quality function $g(f(t))$ and minimizing the playout distortion
function $h(p(t))$. The constraint \eqref{eqn:lyapoptobj1} is the
continuity constraint. Constraints \eqref{eqn:lyapoptobj2} and \eqref{eqn:lyapoptobj3}
are the limits set on $f(t)$ and $p(t)$ respectively.

To see how the above general optimization problem is derived, we first
replace \eqref{eqn:lyapoptobj1} with a continuity constraint derived
from \eqref{eqn:b3}:

\begin{equation}
\mathbb{E}\{F(f(t))-p(t)\}<\beta(t)\label{eqn:opts1}\end{equation}

where $\beta(t)=\frac{b(t)\, r_{min}\, p_{min}}{T-t}$. Constraint
\eqref{eqn:opts1} can be satisfied by either decreasing $f(t)$ and/or
increasing $p(t)$. However, decreasing $f(t)$ would result in a
lower frame quality given by $g(f(t))$ and increasing $p(t)$ would
result in a higher playout distortion given by $h(p(t))$. The optimization
policy would need to handle these tradeoffs. To solve this optimization
problem, we make use of the concepts of virtual buffer and Lyapunov
optimization.

\subsection{Virtual Buffer And Stability\label{sub:Virtual-Buffer}}

Virtual buffers are a concept introduced by Neely et al \cite{ref:neelyvb}
to replace certain constraints of an optimization problem. To determine
a suitable virtual buffer for our problem, we make an initial virtual
buffer design to represent the continuity of the video. The virtual
buffer will be updated at every time slot $t$, and is updated as: 

\begin{equation}
U(t+1)=[U(t)-p(t)]^{+}+F(f(t))\label{eqn:vb1}\end{equation}

where $U(0)=0$. The virtual buffer $U(t)$ is lower bounded by 0
(i.e. always non-negative). This virtual buffer can be seen as the
discontinuity penalty. If $F(f(t))$ is higher than $p(t)$, it means
that the network throughput is lower than the playout rate, i.e. the
rate of video frames received is slower than the amount of video frames
being played out. Thus, the discontinuity penalty $U(t)$ \emph{accumulates}
the difference of $F(f(t))-p(t)$ as a penalty. If the network subsequently
improves and $F(f(t))$ is lower than $p(t)$, the penalty in the
buffer then \emph{reduces} by $p(t)-F(f(t))$. The higher $U(t)$
becomes, the higher the possibility of a buffer underflow. In an ideal
network, where $F(f(t))\leq p(t)$ at all times, $U(t)$ will never
accumulate any penalty.

To ensure that $U(t)$ does not keep increasing, we want $U(t)$ to
\emph{stabilize}. We will show later that by stabilizing $U(t)$,
we can ensure that the video continuity is preserved. The definition
of stability used here is $\mathbb{E}\{U\}\triangleq\limsup_{t\to\infty}\frac{1}{t}\sum_{\tau=0}^{t-1}\mathbb{E}\{U(\tau)\}<\infty$.
This intuitively means that the buffer is stable if it does not grow
infinitely large over time. We also like the virtual buffer to meet
the continuity constraint \eqref{eqn:opts1} when it stabilizes. To
do that, we extend the initial virtual buffer \eqref{eqn:vb1} as: 

\begin{equation}
U(t+1)=[U(t)-p(t)-\beta(t)]^{+}+F(f(t))\label{eqn:vb2}\end{equation}

To see how it works, notice that for the discontinuity penalty $U(t)$
to grow infinitely, the following condition needs to be met: $F(f(t))-p(t)>\beta(t)$.
Therefore, in order for $U(t)$ to stabilize the following condition
needs to be true: $F(f(t))-p(t)\leq\beta(t)$. This is the continuity
constraint as defined in \eqref{eqn:opts1}, so when $U(t)$ stabilizes,
the continuity constraint will be met. The discontinuity penalty $U(t)$
is maintained at the receiver in our design. 

Thus, with the stability of the discontinuity penalty $U(t)$ as a
constraint, we then obtain the general optimization problem presented
in Section \ref{sub:General-Optimization-Problem}. We will demonstrate
in Section \ref{sub:Delay-Constrained-Exp}, using simulations, that
$U(t)$ is positively correlated with the video discontinuity.

\subsection{Lyapunov Optimization Derivation\label{sub:Discontinuity-Penalty-Derivation}}

We show here how we convert the optimization problem presented in
Section \ref{sub:Virtual-Buffer} into a separate encoder and decoder
optimization policies using Lyapunov optimization. We assume that
there is no network delay between the sender and receiver (i.e. $d_{f}=d_{b}=0$).
This is to simplify the analysis presented here. We will relax this
assumption in Section \ref{sec:Network-Delay-Impact}.

We define a Lyapunov function $L(U(t))$ to represent the \textquotedbl{}energy\textquotedbl{}
of the discontinuity penalty $U(t)$ at time $t$, this can be any
arbitrary non-negative function. We use the following Lyapunov function
in this paper: 

\begin{equation}
L(U(t))\triangleq\frac{U^{2}(t)}{2}\label{eqn:lyap}\end{equation}

We then define the one-step conditional Lyapunov drift $\Delta(U(t))$
as: 

\begin{equation}
\Delta(U(t))\triangleq\mathbb{E}\{L(U(t+1))-L(U(t))|U(t)\}\label{eqn:1stepdrift}\end{equation}

Equation \eqref{eqn:1stepdrift} can be understood as the expected
change in the energy in one time slot step. The goal of Lyapunov optimization
is to show that this energy reduces or stays the same at each slot
(i.e. \eqref{eqn:1stepdrift} produces a negative or zero drift).
This would ensure the stability of the buffer, which in turn enforces
the continuity of the video playout. To show that $U(t)$ stabilizes,
we need to convert the buffer update equation \eqref{eqn:vb2} into
a one-step drift \eqref{eqn:1stepdrift}. 

To do that, we square the buffer update equation \eqref{eqn:vb2},
divide it by two, and take expectations of the result (see Section
\ref{sec:apx-drift1-froimp} in the appendix for details),
we will get the following expression: 

\begin{equation}
\Delta(U(t))\leq B-U(t)\mathbb{E}\{\beta(t)+p(t)-F(f(t))|U(t)\}\label{eqn:drift1}\end{equation}

Where $B$ is a constant defined as:

\begin{equation}
B=\frac{1}{2}\bigg(r_{max}^{2}+\bigg(p_{max}+\frac{T\, r_{min}\, p_{min}}{T-t}\bigg)^{2}\bigg)\label{eqn:driftB}\end{equation}

From \eqref{eqn:drift1}, we can use the results in \cite{ref:neelybook}
to prove that $U(t)$ stabilizes (see Section \ref{sub:Discontinuity-Penalty-bnd}
in the appendix). Once $U(t)$ is proven to stabilize,
it can be shown using the results in \cite{ref:neelyvb} that the
continuity constraint \eqref{eqn:opts1} is satisfied. However, to
optimize the frame quality utility $g(f(t))$ and the playout distortion
$h(p(t))$, we need to massage the equation more. By subtracting from
both sides, the term $V\mathbb{E}\{g(f(t))-h(p(t))|U(t)\}$, which
is the expectation of \eqref{eqn:lyapoptobj} scaled by a positive
constant $V>0$, and by rearranging the terms. We get: 

\begin{align}
\Delta(U(t)) & -V\mathbb{E}\{g(f(t))-h(p(t))|U(t)\} & \notag\label{eqn:lyapobj}\\
\leq B & -\mathbb{E}\{U(t)\beta(t)|U(t)\} & \notag\\
 & -\mathbb{E}\{Vg(f(t))-U(t)F(f(t))|U(t)\} & \notag\\
 & -\mathbb{E}\{U(t)p(t)-Vh(p(t))|U(t)\}\end{align}

The third term on the right hand side of \eqref{eqn:lyapobj} is a
function of the encoder frame generation interval $f(t)$ only and
represents the sender optimization policy. The last term of \eqref{eqn:lyapobj}
is a function of the playout frame interval $p(t)$ and represents
the receiver optimization policy.

To summarize:

\subsubsection{Encoder optimization policy}

From the third term of \eqref{eqn:lyapobj}, based on the frame interval
scaling factor $e(t)$ (defined in \eqref{eqn:delayscale}) and discontinuity
penalty $U(t)$ feedback from the receiver. The encoder in the sender
will calculate $F(f(t))$ using \eqref{eqn:Fdefine}, and it will
choose $f(t)$ at each time slot as the solution of the following
optimization: 

\begin{align}
\textrm{Maximize:} & \quad Vg(f(t))-U(t)F(f(t))\notag\label{eqn:encobj}\\
\textrm{Subject to:} & \quad f_{min}\leq f(t)\leq f_{max}\end{align}

\subsubsection{Decoder optimization policy }

From the last term of \eqref{eqn:lyapobj}, the decoder in the receiver
will observe the current discontinuity penalty $U(t)$ and choose
$p(t)$ at each time slot as the solution of the following optimization: 

\begin{align}
\textrm{Maximize:} & \quad U(t)p(t)-Vh(p(t))\notag\label{eqn:decobj}\\
\textrm{Subject to:} & \quad p_{min}\leq p(t)\leq p_{max}\end{align}

Notice that the optimization policies are decoupled into separate
optimization subproblems for playout frame and encoder frame generation
intervals. Note that the decoder is responsible for updating $U(t)$
using equation \eqref{eqn:vb1} and sending the value of $U(t)$ to
the encoder. Furthermore, under appropriate conditions, the decoupled
problems are convex. This makes the problem easier and more flexible
to solve. Given that Lyapunov optimization minimizes the right hand
side of \eqref{eqn:lyapobj} instead of the original optimization
problem \eqref{eqn:lyapoptobj} (which is hard to solve), the objective
function value realized by Laypunov optimization is sub-optimal but
its deviation from the optimal value can be controlled (see Section
\ref{sub:Discontinuity-Penalty-bnd} in the appendix).

\section{Delay Constrained Frame Rate Optimization\label{sec:Delay-Constrained-Frame}}

While slowing down the video playout allows us to reduce the occurrences
of buffer underflows and preserve the video continuity, it introduces
extra viewing latency to the viewers. This becomes an issue when the
system has a delay budget, as frequent playout slowdowns might cause
the delay budget to be exceeded.

In this section, we examine the problem when there is a constraint
imposed on the viewing latency. We focus on constraining the additional
playout latency generated by slowing down the playout. More specifically,
we want to impose a constraint on how often playout slowdowns occur
and how much the playout can be slowed down. This is done by introducing
another virtual buffer, called the \emph{delay accumulator}, into
the problem to represent the constraint on the accumulated delay due
to playout slowdowns. We then use Lyapunov optimization to ensure
that this constraint is met.

\subsection{Delay Constrained Policy Design}

The delay constraint can be described as constraining the accumulated
playout slowdowns used over the lifetime of the whole video sequence.
Specifically, let $\theta$ be the maximum playout slowdown delay
tolerable for the video application and $p_{n}$ represent the natural
playout interval of the video. Then, the constraint could be written
as: 

\begin{equation}
\sum_{\tau=0}^{T-1}(p(\tau)-p_{n})\leq\theta\label{eqn:delcon}\end{equation}

Recall that $T$ is the length of the sequence in time slots. Since
Lyapunov optimization works on a per time slot basis, we need to express
the constraint \eqref{eqn:delcon} as a constraint for each time slot.
To do this, note that for the current timeslot $t$, the averaged
maximum playout slowdown delay tolerable will be: $t_{d}=\frac{\theta}{T-t}$,
where $T-t$ represents the number of time slots remaining for the
sequence. With this, we can rewrite the delay constraint \eqref{eqn:delcon}
as:

\begin{equation}
\mathbb{E}\{p(t)-p_{n}\}\leq t_{d}\label{eqn:delcon2}\end{equation}

Then, what remains to be done is to design a policy that ensures that
constraint \eqref{eqn:delcon2} is met at each time slot. It can be
seen that by adding constraint \eqref{eqn:delcon2} to the general
optimization problem presented in Section \ref{sub:Virtual-Buffer},
we obtain a problem that provides video continuity while ensuring
that the delay constraint is met.

\subsection{Delay Accumulator Virtual Buffer}

To apply the delay constraint \eqref{eqn:delcon2} into the framework
using Lyapunov optimization, we again make use of the virtual buffer
concept. We introduce another virtual buffer named as the \emph{delay
accumulator}. The delay accumulator is a virtual buffer that keeps
track of the accumulated delay caused by playout slowdowns. Everytime
a playout slowdown is perform, the delay accumulator will increase
correspondingly. However, the delay accumulator will reduce when playout
speed up is perform. Formally, the buffer dynamics to describe the
delay accumulator for each time slot would be:

\begin{equation}
X(t+1)=[X(t)-p_{n}-t_{d}]^{+}+p(t)\label{eqn:Xbuff}\end{equation}

Note that the terms $p_{n}$, $p(t)$ and $t_{d}$ are taken from
constraint \eqref{eqn:delcon2} above. Therefore, by showing that
the delay accumulator $X(t)$ stabilizes, we can show that the delay
constraint \eqref{eqn:delcon2} can be satisfied \cite{ref:neelyvb}.
This also means that the delay constraint can be written as: $X(t)\textrm{ is stable}$.
We now show how Lyapunov optimization can be used to derive optimization
policies to solve the above problem.

\subsection{Lyapunov Optimization Derivation}

Note that there are two buffers in the problem, the discontinuity
penalty $U(t)$ and the delay accumulator $X(t)$. To stabilize both
of these simultaneously, we first redefine the Lyapunov function to
be:

\begin{equation}
L(U(t),X(t))\triangleq\frac{U^{2}(t)+X^{2}(t)}{2}\label{eqn:Xlyap}\end{equation}

The one step conditional drift also needs to consider both buffers
and is defined as:

\begin{align}
\Delta(U(t),X(t))\triangleq\; & \mathbb{E}\{L(U(t+1),X(t+1))\notag\label{eqn:Xonestep}\\
 & \quad-L(U(t),X(t))|U(t),X(t)\}\end{align}

To shorten the formulas, we use $\Delta$, $U$, $X$, $p$ and $f$
to represent $\Delta(U(t),X(t))$, $U(t)$, $X(t)$, $p(t)$ and $f(t)$
respectively. By squaring \eqref{eqn:vb2} and \eqref{eqn:Xbuff},
taking expectations and dividing by 2, we get (see Section \ref{sec:apx-Xdrift1-froimp}
in the appendix for details): 

\begin{align}
\Delta(U,X)\leq & \; B+C-U\mathbb{E}\{p-F(f)|U,X\}\notag\label{eqn:Xdrift1}\\
 & \quad-X\mathbb{E}\{p_{n}+t_{d}-p|U,X\}\end{align}

where $B$ is defined as in \eqref{eqn:driftB} and $C=\frac{1}{2}\bigg(p_{max}^{2}+(p_{n}+t_{d})^{2}\bigg)$.

It can be proven that \eqref{eqn:Xdrift1} results in stability for
both the discontinuity $U$ and the delay accumulator $X$ \cite{ref:neelyvb}
(see Section \ref{sub:Delay-Constrained-bnd} of the supplementary
materials). Furthermore, it can be proven that the stabilization of
$X$ implies that the constraint \eqref{eqn:delcon2} can be met by
using the results from \cite{ref:neelyvb}. 

To optimize the frame quality and the playout distortion, we subtract
$V\mathbb{E}\{g(f(t))-h(p(t))|U(t)\}$ from both sides of \eqref{eqn:Xdrift1}
to obtain: 

\begin{align}
\Delta(U,X) & -V\mathbb{E}\{g(f)-h(p)|U,X\} & \notag\label{eqn:Xlyapobj}\\
\leq B+C & -X\mathbb{E}\{p_{n}+t_{d}|U,X\}\notag\\
 & -\mathbb{E}\{Vg(f)-UF(f)|U,X\} & \notag\\
 & -\mathbb{E}\{Up-Vh(p)-Xp|U,X\}\end{align}

Note that the encoder optimization policy (second last term) remains
the same as \eqref{eqn:encobj}. The last term shows that decoder
optimization policy \eqref{eqn:decobj} has an additional penalty
term $-Xp$. This will mean that as $X$ increases, the decoder will
get penalized more for high playout interval $p$ values. This will
encourage the decoder to choose a lower $p$ whenever the accumulated
delay in $X$ is high. Lastly, the performance bound for the Lyapunov
optimization can be derived (see Section \ref{sub:Delay-Constrained-bnd}
of the appendix).

\section{Network Delay Impact\label{sec:Network-Delay-Impact}}

In Sections \ref{sec:Discontinuity-Penalty} and \ref{sec:Delay-Constrained-Frame}.
we showed how Lyapunov optimization can derive optimization policies
that help ensure the video continuity and enforce a delay constraint.
However, the assumption made in those sections for the derivations
is that there is no network delay. Specifically, we assumed that the
feedback delay from the receiver to the sender is non-existent. In
this section, we relax this network delay assumption and analyze the
impact of network delay on the optimization policies.

Recall that the network generates a forward delay of $d_{f}$ and
a backward delay of $d_{b}$ from the system model discussed in Section
\ref{sub:fro-imp-System-Model}. With network delay, the discontinuity
penalty $U(t)$ updating in the receiver is performed as: 

\begin{equation}
U(t+1)=[U(t)-\gamma(t)]^{+}+F(f(t-d_{f}))\label{eqn:rxbuf}\end{equation}

where $\gamma(t)=p(t)+\beta(t)$ and recall that $\beta(t)=\frac{b(t)\, r_{min}\, p_{min}}{T-t}$.
Note that the difference between the previously presented discontinuity
penalty buffer dynamics in \eqref{eqn:vb1} and above is that \eqref{eqn:rxbuf}
is based on the forward delayed encoder frame generation interval
$f(t-d_{f})$. Furthermore, the sender relies on feedback from the
receiver, so it chooses $f(t)$ based on a backward delayed $U(t-d_{b})$.
There are two possible issues that might impact on the optimization
policies when delays are present: 
\begin{enumerate}
\item the delayed discontinuity penalty $U(t)$ feedback from the receiver
to the sender means the encoder optimization policy will need to make
use of $U(t-d_{b})$. 
\item \label{enu:fro-imp_issue-2}the current choice of $f(t)$ at the sender
would only affect receiver $d_{f}$ time slots later.
\end{enumerate}
What we need to do is to derive optimization policies that take the
above two issues into account and show that these policies stabilizes
the discontinuity penalty $U(t)$.

To begin, note that \eqref{eqn:rxbuf} can be represented recursively
as: 

\begin{align}
U(t+1) & =[U(t)-\gamma(t)]^{+}+F(f(t-d_{f}))\notag\\
U(t) & =[U(t-1)-\gamma(t-1)]^{+}+F(f(t-d_{f}-1))\notag\\
 & \vdots\notag\\
U(t-d_{b}+1) & =[U(t-d_{b})-\gamma(t-d_{b})]^{+}+F(f(t-d_{f}-d_{b}))\end{align}

This implies that $U(t)$ can be recursively defined as: 

\begin{equation}
U(t)\leq\left[U(t-d_{b})-\sum_{\tau=t-d_{b}}^{t-1}\gamma(\tau)\right]^{+}+\sum_{\tau=t-d_{b}-d_{f}}^{t-d_{f}-1}F(f(\tau))\label{eqn:ut}\end{equation}

Issue 2 suggests that we need to predict $d_{f}$ time slots in the
future. To do that, we change the buffer updating equation \eqref{eqn:rxbuf}
into a $d_{f}$ slot update:

\begin{equation}
U(t+d_{f}+1)\leq\left[U(t)-\sum_{\tau=t}^{t+d_{f}}\gamma(\tau)\right]^{+}+\sum_{\tau=t-d_{f}}^{t}F(f(\tau))\label{eqn:currfut}\end{equation}

What \eqref{eqn:currfut} means is that the receiver updates the discontinuity
penalty $U(t)$ not only based on the known current values $\gamma(t)$
and $f(t-d_{f})$ but also using the predicted values $d_{f}$ slots
in the future, $\left[\gamma(t+1)\ldots\gamma(t+d_{f})\right]$ and
$\left[f(t-d_{f}+1)\ldots f(t)\right]$. Note that $F(f(\tau))$ in
\eqref{eqn:currfut} is calculated from $d_{f}$ time steps in the
\emph{past}. This is because the current $f(t)$ only affects the
discontinuity penalty $d_{f}$ time slots later which will be $U(t+d_{f}+1)$.
The $d_{f}$ step buffer dynamics can be proved to obtain stability
by using the \emph{T-slot Lyapunov drift} \cite{ref:neelybook}. 

To show how the T-slot Lyapunov drift achieves stability, we first
convert the 1-step Lyapunov drift in \eqref{eqn:1stepdrift} into
a $d_{f}$-step drift:

\begin{equation}
\Delta(U(t))\triangleq\mathbb{E}\{L(U(t+d_{f}+1))-L(U(t))|U(t)\}\label{eqn:dfstepdrift}\end{equation}

We now use the following shortened notations to simplify the equations:
$U\triangleq U(t)$, $U_{d_{b}}\triangleq U(t-d_{b})$, $\gamma\triangleq\sum_{\tau=t-d_{b}}^{t-1}\gamma(\tau)$,
$\gamma_{d_{f}}\triangleq\sum_{\tau=t}^{t+d_{f}}\gamma(\tau)$, $F\triangleq\sum_{\tau=t-d_{b}-d_{f}}^{t-d_{f}-1}F(f(\tau))$
and $F_{d_{f}}\triangleq\sum_{\tau=t-d_{f}}^{t}F(f(\tau))$.

Using the same Lyapunov function \eqref{eqn:lyap} as in the previous
section and drift definition \eqref{eqn:dfstepdrift}. If we square
\eqref{eqn:currfut}, divide it by two and take expectations (see
Section \ref{sec:apx-deldrift1-froimp} of the appendix
for details), we get: 

\begin{equation}
\Delta(U)\leq B'-\mathbb{E}\left\{ U\gamma_{d_{f}}\big|U\right\} +\mathbb{E}\left\{ UF_{d_{f}}\big|U\right\} \label{eqn:deldrift1}\end{equation}

Where: 

\begin{equation}
B'=\frac{d_{f}}{2}\bigg(r_{max}^{2}+\bigg(p_{max}+\frac{T\, r_{min}\, p_{min}}{T-t}\bigg)^{2}\bigg)\label{eqn:driftBdelay}\end{equation}

Equation \eqref{eqn:deldrift1} can be shown to achieve stability
\cite{ref:neelythesis}, effectively settling issue 2. However, to
deal with issue 1, we need to show how the feedbacked discontinuity
penalty $U(t-d_{b})$ affects \eqref{eqn:deldrift1}. To do that,
we substitute the recursively defined $U(t)$ \eqref{eqn:ut} into
\eqref{eqn:deldrift1}: 

\begin{equation}
\Delta(U)\leq\; B'-\mathbb{E}\left\{ U\gamma_{d_{f}}\big|U\right\} +\mathbb{E}\left\{ \left(\left[U_{d_{b}}-\gamma\right]^{+}+F\right)F_{d_{f}}\big|U\right\} \label{eqn:drift3}\end{equation}

Recall the definitions of $F$ and $F_{d}{}_{f}$, and given that
$F(.)$ is an increasing function of $f(t)$ (see \eqref{eqn:Fdefine}),
this implies that $F$ and $F_{d_{f}}$ can be bounded as $F\leq d_{b}r_{max}$
and $F_{d_{f}}\leq(d_{f}+1)r_{max}$ respectively. It then follows
that $FF_{df}$ can be bounded as:

\begin{equation}
FF_{d_{f}}\leq d_{b}(d_{f}+1)r_{max}^{2}\label{eqn:sumbound}\end{equation}

Note that $\left[U_{d_{b}}-\gamma\right]^{+}\leq U_{d_{b}}$. Thus,
we use \eqref{eqn:sumbound} in \eqref{eqn:drift3}, we get: 

\begin{equation}
\Delta(U)\leq\; B''-U\mathbb{E}\left\{ \gamma_{d_{f}}\big|U\right\} -\mathbb{E}\left\{ -U_{d_{b}}F_{d_{f}}\big|U\right\} \label{eqn:drift4}\end{equation}

where $B''=B'+d_{b}(d_{f}+1)\; r_{max}^{2}$ with $B'$ from \eqref{eqn:driftBdelay}.
Equation \eqref{eqn:drift4} in that form can be proven to stabilize
by using the results from \cite{ref:neelythesis} (see Section \ref{sub:Perf-Bnds-del}
in the appendix), thus settling issue 1. 

As in Section \ref{sub:Discontinuity-Penalty-Derivation}, to cater
for utility optimization, we subtract from both sides, the term $V\mathbb{E}\{g(f(t))-h(p(t))|U\}$
and rearrange the terms to obtain:

\begin{align}
\notag\Delta(U) & -V\mathbb{E}\{g(f(t))-h(p(t))|U\}\\
\notag & \leq\; B''-\mathbb{E}\left\{ U\gamma_{d_{f}}-Vh(p(t))\big|U\right\} \\
 & \quad\quad\quad-\mathbb{E}\left\{ Vg(f(t))-U_{d_{b}}F_{d_{f}}\big|U\right\} \label{eqn:delcon-deriv1}\end{align}

Using the definitions of $\gamma_{d_{f}}$ and $F_{d_{f}}$ and that
$\gamma(t)=p(t)+\beta(t)$. Equation \eqref{eqn:delcon-deriv1} can
be rewritten as:

\begin{align}
\notag\Delta(U) & -V\mathbb{E}\{g(f(t))-h(p(t))|U\}\\
\notag & \leq\; B''-U\mathbb{E}\left\{ \sum_{\tau=t+1}^{t+d_{f}}p(\tau)+\sum_{\tau=t}^{t+d_{f}}\beta(\tau)\Bigg\vert U\right\} \\
\notag & \quad\quad\quad+U_{d_{b}}\mathbb{E}\left\{ \sum_{\tau=t-d_{f}}^{t-1}F(f(\tau))\Bigg\vert U\right\} \\
\notag & \quad\quad\quad-\mathbb{E}\left\{ Up(t)-Vh(p(t))\vert U\right\} \\
 & \quad\quad\quad-\mathbb{E}\left\{ Vg(f(t))-U_{d_{b}}F(f(t))\vert U\right\} \label{eqn:delcon-deriv2}\end{align}

It can be seen from \eqref{eqn:delcon-deriv2} that the last two terms
represent the decoder and encoder optimization policies respectively.
The decoder policy is not affected by the network delay while the
only change to the encoder policy is to make use of $U(t-d_{b})$
fed back from the decoder. Moreover, since the delay accumulator $X(t)$
is updated locally within the decoder. This would imply that the delay
constrained decoder policy \eqref{eqn:Xlyapobj} would not be affected.
Lastly, the performance bound for the Lyapunov optimization can be
derived (see Section \ref{sub:Perf-Bnds-del} of the supplementary
materials).

\section{\label{sec:Video-Quality-Functions}Video Quality Functions}

In the previous sections, we showed how using the discontinuity penalty
virtual buffer in the system model allows us to express the processes
as frame intervals. We then used Lyapunov optimization to derive optimization
policies that stabilizes the discontinuity penalty virtual buffer
and showed that this helps to maintain the video continuity. We also
demonstrated how to add a delay constraint into a framework by using
the delay accumulator virtual buffer. By deriving optimization policies
that stabilze the delay accumulator virtual buffer, we showed that
the delay constraint can be met. Finally, we studied the impact of
network delay on the optimization policies and showed how the optimization
policies can be derived with network delay consideration.

What is lacking thus far is a discussion on the specific choice of
frame quality function $g(f(t))$ and playout distortion function
$h(p(t))$. In this section, we shall examine the specific forms for
$g(f(t))$ and $h(p(t))$.

\subsection{Frame Quality Function\label{sub:Frame-Quality-Function}}

We first look at an appropriate frame quality function for $g(f(t))$,
$g(f(t))$ should ideally be concave so that a solution can be easily
found for the encoder policy \eqref{eqn:encobj}. As mentioned before
as $f(t)$ decreases, the encoder frame generation rate increases.
This means that more compression is needed to meet the available network
bandwidth and more compression tends to mean that the frame quality
will be reduced. One of the ways to measure frame quality is to measure
the peak signal-to-noise-ratio (PSNR) of the video.

\begin{figure}[tbh]
\begin{minipage}[t]{0.45\textwidth}%
\includegraphics[scale=0.3]{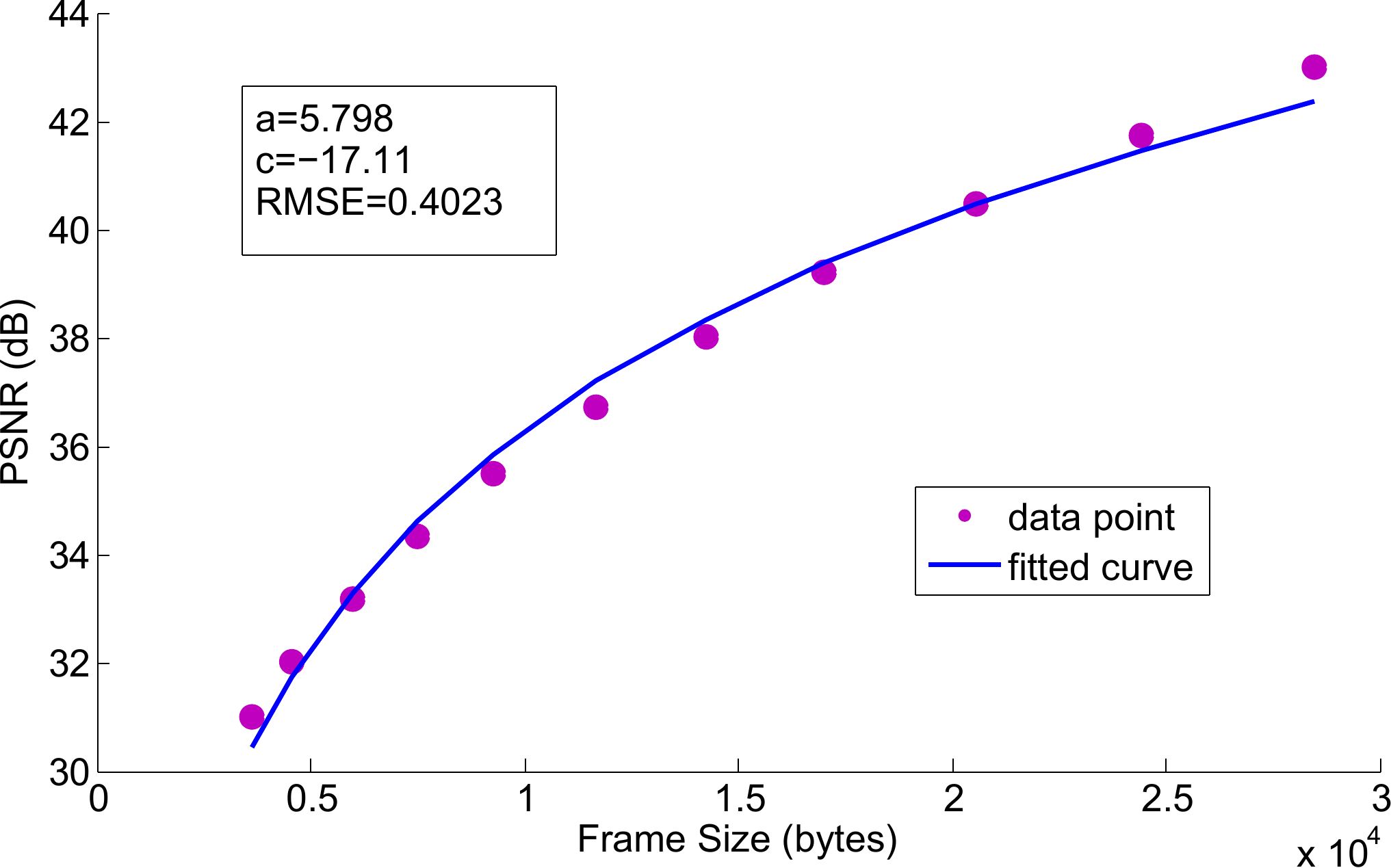}

\caption{PSNR versus average frame size.}

\label{fig:fr_psnr}%
\end{minipage}\hfill{}%
\begin{minipage}[t]{0.45\textwidth}%
\includegraphics[scale=0.35]{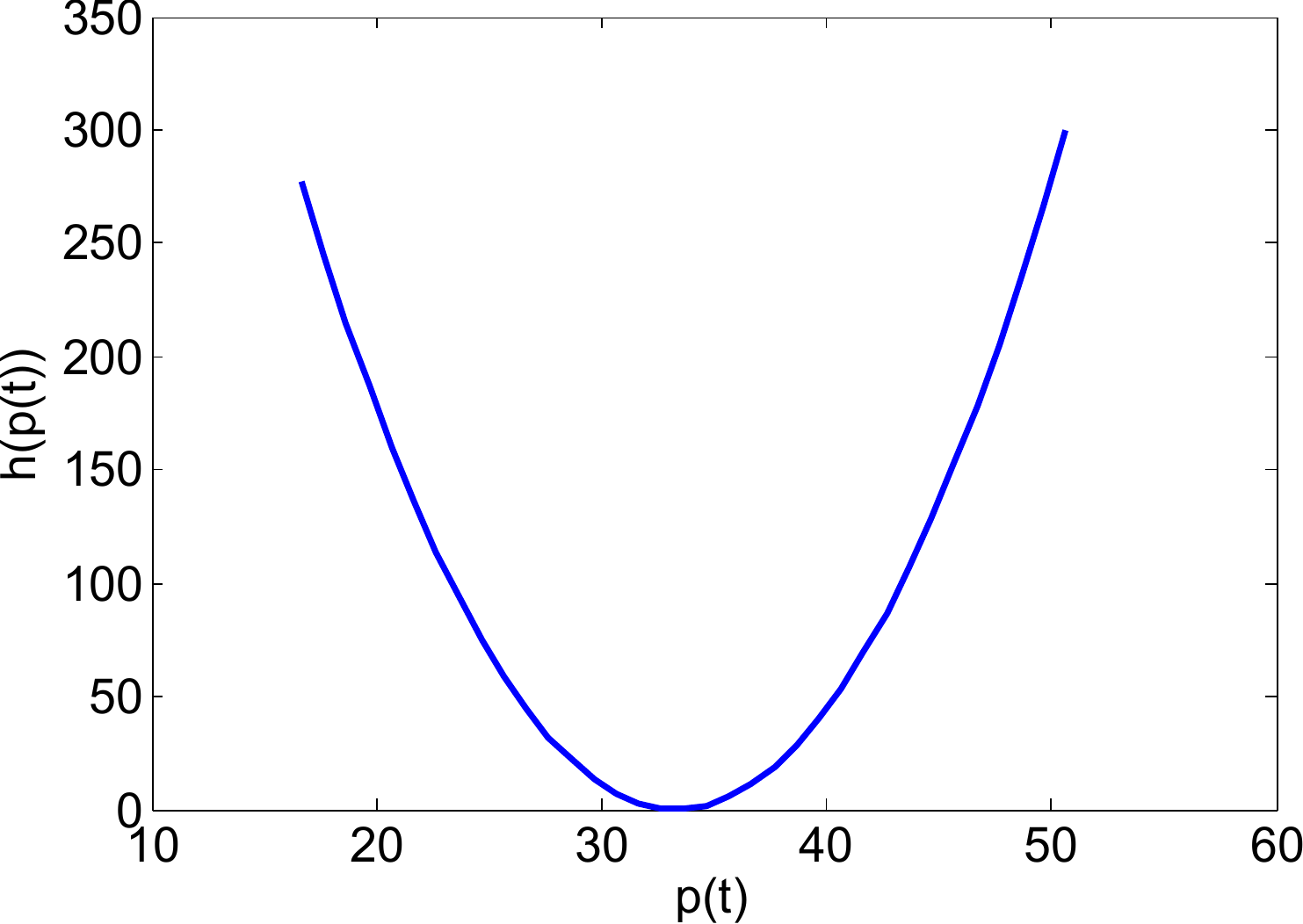}

\caption{Playout distortion function. $p(t)$ is in milliseconds, $m=1$ and
$p_{n}=\frac{1}{30}$ ms.}

\label{fig:podist}%
\end{minipage}
\end{figure}

An and Nguyen \cite{ref:vidutilities} has been shown that PSNR can
be represented using a log function of the bitrate. We make use of
their result and fit PSNR to the average frame size of the sequence: 

\begin{equation}
PSNR(f(t))=a\, log(ABR(t)\, f(t))+c\label{eqn:frmpsnr}\end{equation}

where $a$ and $c$ are the modeling coefficients, and $ABR(t)$ is
the current available network bandwidth. Note that we make use of
all the available bandwidth to transmit the video frames from the
sender. Fig. \ref{fig:fr_psnr} shows the fitted curve. The video
sequence used for fitting is a concatenation of football, city, crew
and akiyo sequences in that order. This is done to ensure that the
sequence contains several subsequences of different coding complexity.
Notice that if we use the encoder frame generation rate $i(t)$ as
the input variable instead, \eqref{eqn:frmpsnr} becomes: 

\begin{equation}
PSNR(i(t))=a\, log\bigg(\frac{ABR(t)}{i(t)}\bigg)+c\label{eqn:invfrmpsnr}\end{equation}

where $i(t)=\frac{1}{f(t)}$, the resulting PSNR function would no
longer be a concave function of $i(t)$ and would make the optimization
problem more difficult to solve. This is the reason why we use the
encoder frame generation interval $f(t)$ instead of encoder frame
generation rate $i(t)$.

Since $f(t)$ is upper bounded by $f_{max}$, we fix the maximum of
$PSNR(f(t))$ to $f_{max}$. This is done to make the PSNR function
differentiable in the interval $[f_{min},f_{max}]$ and make the maximization
of it easy to calculate. We do this by subtracting $\frac{a}{f_{max}}$
from the first derivative of \eqref{eqn:frmpsnr} ($PSNR'(f(t))$): 

\begin{equation}
PSNR'(f(t))-\frac{a}{f_{max}}=\frac{a}{f(t)}-\frac{a}{f_{max}}\label{eqn:frmpsnrmax}\end{equation}

Integrating \eqref{eqn:frmpsnrmax} will give us the desired $g(f(t))$:

\begin{equation}
g(f(t))=a\, log(ABR(t)\, f(t))+c-\frac{a\, f(t)}{f_{max}}\label{eqn:g}\end{equation}

Note that $g(f(t))$ is concave between the range $[f_{min},f_{max}]$
(where $f_{min}>0$).

\subsection{Playout Distortion Function}

In this section, we choose an appropriate playout distortion function
for $h(p(t))$. We modified the version of playout distortion function
used in \cite{ref:ampscheduler5}: 

\begin{equation}
h(p(t))=m\cdot(p_{n}-p(t))^{2}\label{eqn:h}\end{equation}

where $m$ is the motion intensity of the sequence, calculated using
the technique in \cite{ref:frquality}, and $p_{n}$ is the natural
playout interval. Fig. \ref{fig:podist} shows the playout distortion
function $h(p(t))$. $h(p(t))$ is convex in the range $[p_{min},p_{max}]$.
Eqn. \eqref{eqn:h} is a combination of the quadratic playout distortion
function proposed in \cite{ref:ampscheduler1,ref:quadplayout} and
the motion scaling used in \cite{ref:ampscheduler5}. The idea is
that the playout distortion increases as the playout rate deviates
from the natural playout rate. The playout distortion is also affected
by the motion intensity of the sequence. Intuitively, higher motion
sequences increases playout distortion more as the change in motion
is more perceivable when the playout rate deviates from the natural
playout rate.

\section{\label{sec:Performance-Evaluation}Performance Evaluation}

\subsection{Experiment Setup}

We made use of ns-2 \cite{ref:ns2} to simulate a network with time-varying
data bandwidths. We implemented our framework into a x264 encoder
\cite{ref:x264}, an open source multi-threaded H.264/AVC encoder.
Our implementation in x264 simulates the network transmission using
network traces. The decoder buffer evolution is simulated by tracking
the arrivals of video frames from the simulated network and the removal
of video frames from the buffer for playout. Everytime a frame is
encoded, the framework will make a decision on the encoder frame generation
rate and the playout frame rate at the simulated decoder. This is
to simulate the real-time adjustment of parameters as the video is
being encoded for transmission. Network traces obtained from ns-2
are used to simulate the network within the encoder. The ns-2 traces
provide the sending and receiving times as well as the loss status
of each video packet. Every packet produced by x264 is assigned a
sending time, receiving time and a loss status. These infomation is
used to simulate the sender and receiver buffers at a packet level.
The video packets that arrived at the receiver buffer are then used
to calculate the amount of frames that the decoder buffer contains.
A decoder buffer underflow occurs when the frames removed from the
decoder buffer exceeds the number of frames in it. Note that our framework
could easily be adapted to multiple pre-encoded copies of the video.

For the network simulations, we made use of a dumbell network topology
with the bottleneck bandwidth set to 5 Mbits/s. One pair of nodes
in the network simulated video streaming using the TFRC transport
protocol \cite{ref:tfrc}. To generate background traffic, random
webpage sessions were used for the other pairs of nodes. All the random
sessions go through the bottleneck link. The average packet delay
from the encoder to the decoder is 235 ms, while the feedback delay
from the decoder to encoder is 330 ms. These high delay values enable
us to show that the our proposed distributed frame rate control algorithm
works in presence of delay.

The encoder receives feedback from the decoder and solves the optimization
problem to determine the encoder frame generation interval $f(t)$.
The x264 encoder makes use of this encoder frame generation interval
$f(t)$ by setting the target frame rate of the rate controller to
encoder frame generation rate $\frac{1}{f(t)}$. The rate controller's
target frame rate is used to compute the quantization level for each
frame. The lower the target frame rate, the lower the quantization
level and, as a result, bigger frame sizes. More details on x264's
rate control can be found in \cite{ref:x264rc}.

The test sequence used is a concatenation of football, city, crew
and akiyo sequences in that order. This is to ensure a mix of high
and low motion within the sequence. A 16 minute test sequence is obtained
by repeating the concatenated sequence.

The model coefficient $a$ from \eqref{eqn:g} is found by curve fitting
to be 4.91. The constant $V$ is set to 1. In our experiments, we
tested the continuity of the video, this is defined here as the amount
of time spent playing video, specifically: 

\begin{equation}
\text{Playout Continuity}=1-\frac{\text{rebuffering time}}{\text{total sequence time}}\label{eqn:cont}\end{equation}

where the rebuffering time is the time needed to refill the buffer
to a certain threshold after an underflow event, this is a typical
behaviour of a video player \cite{ref:rebuff}. The rebuffering threshold
is set to half of the prebuffer amount. The prebuffer amount is varied
for each run to determine the performance of each scheme. At the end
of each run, we will calculate the playout continuity using \eqref{eqn:cont}
for each scheme and make a comparison.

\subsection{Discontinuity Penalty Lyapunov Optimization Results\label{sub:Discontinuity-Penalty-Exp}}

\begin{figure}[tbh]
\begin{minipage}[t]{0.45\textwidth}%
\includegraphics[scale=0.5]{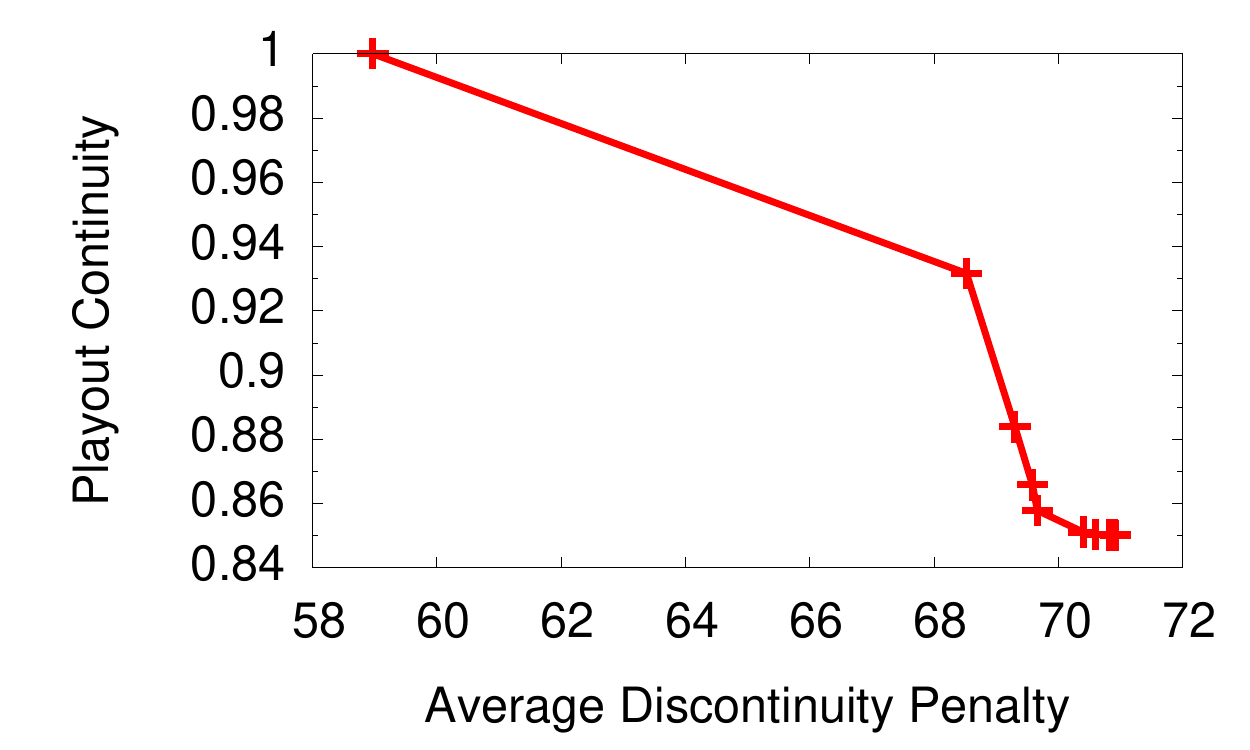}

\caption{Correlation between average discontinuity penalty and continuity.}

\label{fig:ucont}%
\end{minipage}\hfill{}%
\begin{minipage}[t]{0.45\textwidth}%
\includegraphics[scale=0.5]{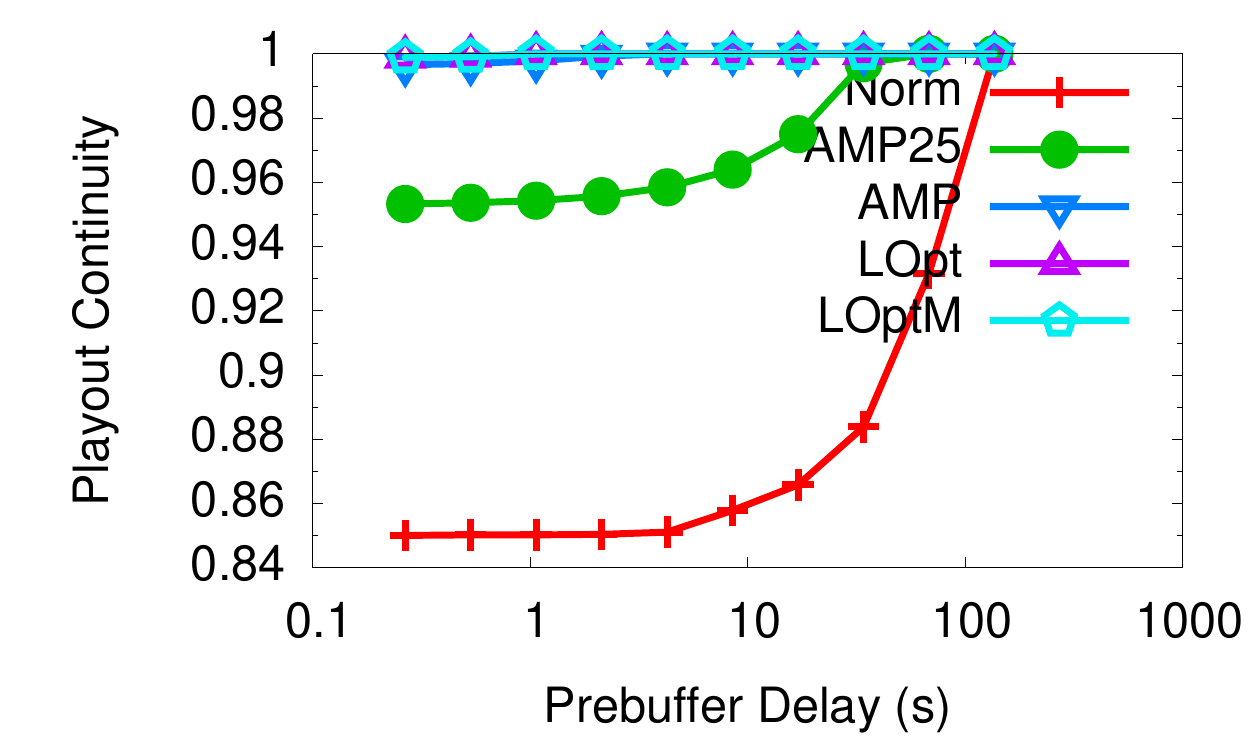}

\caption{Prebuffering delay vs continuity. }
\label{fig:prerollcont}%
\end{minipage}\hfill{}%
\begin{minipage}[t]{0.45\textwidth}%
\includegraphics[scale=0.5]{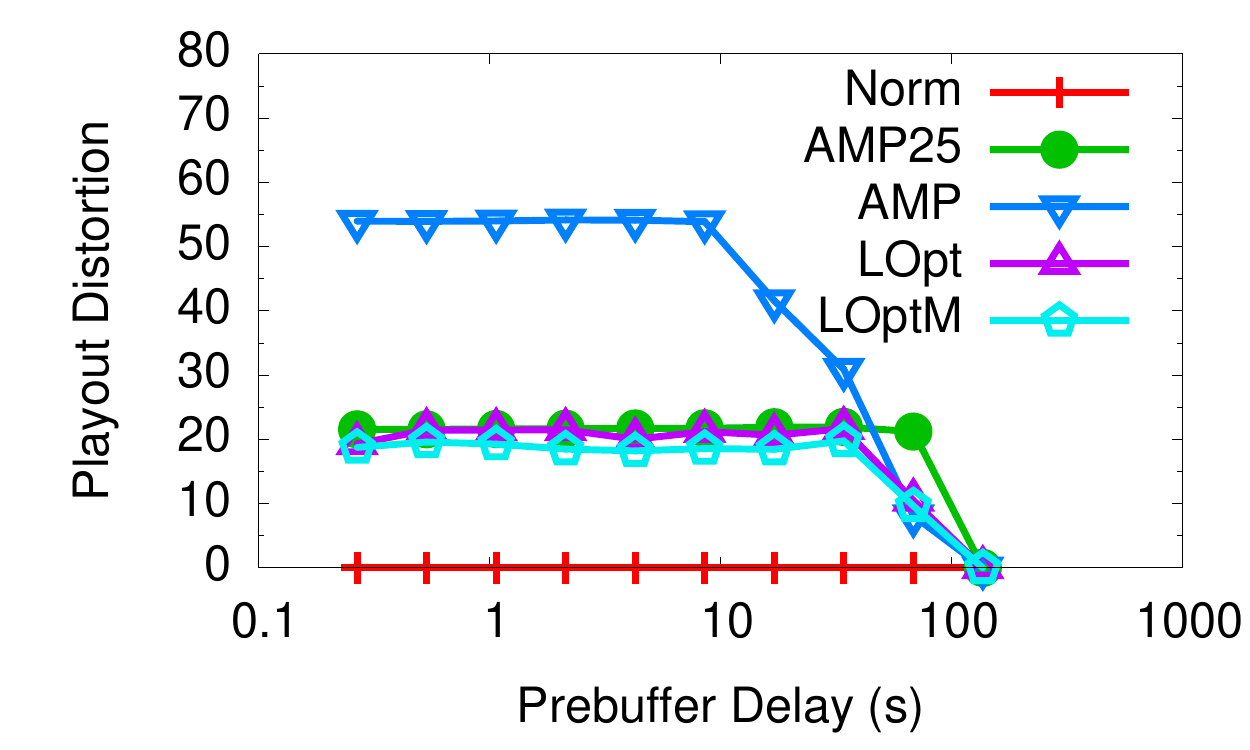}

\caption{Prebuffering delay vs playout distortion. }
\label{fig:prerolldist}%
\end{minipage}\hfill{}%
\begin{minipage}[t]{0.45\textwidth}%
\includegraphics[scale=0.5]{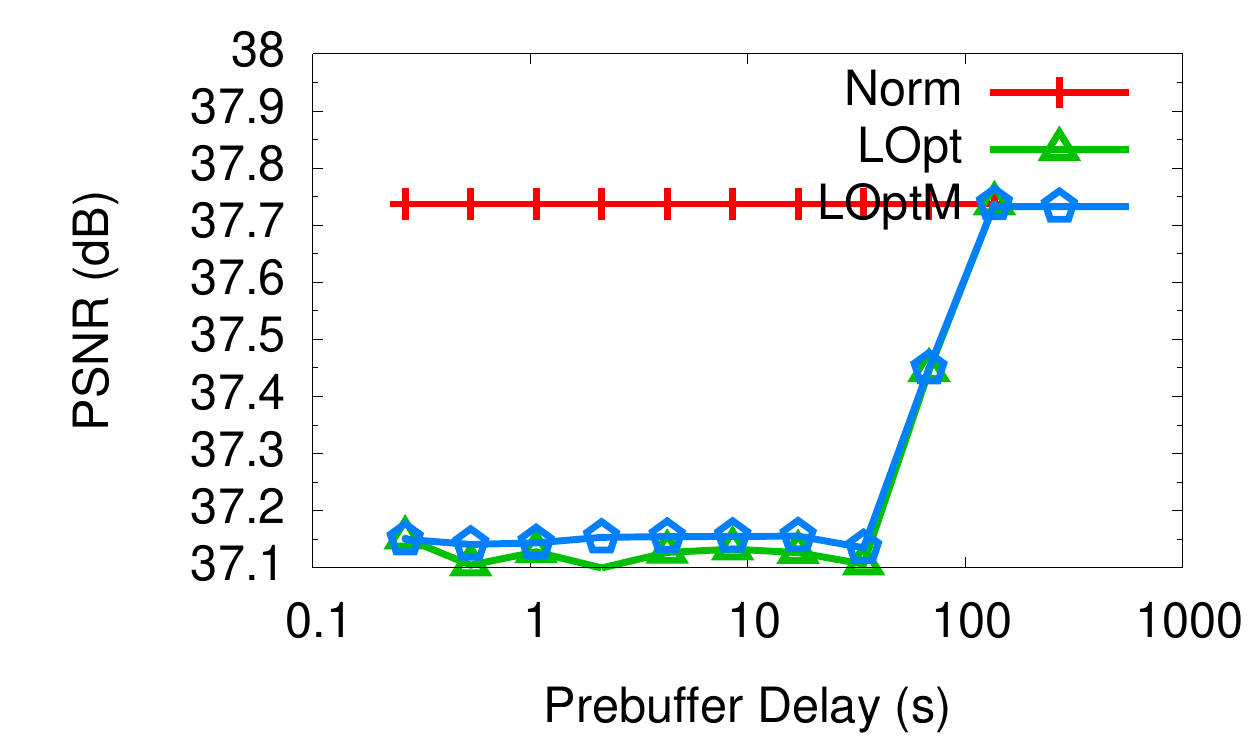}

\caption{PSNR loss for \emph{LOpt}.}
\label{fig:prerollpsnr}%
\end{minipage}
\end{figure}

We first examine the (negative) correlation between the discontinuity
penalty $U(t)$ and the continuity of the video \eqref{eqn:cont},
fig. \ref{fig:ucont} plots this. It can be seen from the graph that
as the discontinuity penalty increases, the continuity drops due to
the longer rebuffering time caused by an increased number of buffer
underflows. This shows that the discontinuity penalty $U(t)$ can
be used as an indicator of a lack of video continuity.

We next present the results of the Lyapunov optimization framework
described in Section \ref{sec:Discontinuity-Penalty}, labelled as
\emph{LOpt} here. We also set up our framework with a less conservative
bound than the one described in \eqref{eqn:b3}, this is labelled
as \emph{LOptM}. This is done by setting $\beta(t)=\frac{b(t)\, r_{max}\, p_{max}}{T-t}$
in \eqref{eqn:vb2}. We compare our framework with a typical setup
of x264 with its target frame rate and its playout rate set to a constant
30 fps, we label this scheme \emph{Norm}. We also compared our framework
with the AMP scheme \cite{ref:ampkalman}. We implemented a combination
of AMP-Initial and AMP-Robust. AMP-Initial slows down the playout
rate by a predetermined slowdown factor when the video starts playing
until a certain buffer level has been reached. AMP-Robust slows down
the playout rate of the video by a predetermined slowdown factor when
the buffer level falls below a certain level. In our implementation,
AMP-Initial is used in conjunction with AMP-Robust. We empirically
chose the smallest slowdown factor of AMP such that it achieves a
continuity of 99\% or higher in each test case. This effectively simulates
a close to optimal adaptive AMP scheme, we label this as \emph{AMP}.
We also added in the results of \emph{AMP} with a constant slowdown
factor of 25\% (used in \cite{ref:ampkalman}) as a reference, which
we label \emph{AMP25}.

To examine the performance of each scheme, we compare the continuity
of each scheme based on the amount of prebuffering provided. Note
that in the results, the prebuffering delay is on a log base 10 scale.
Continuity is calculated as in \eqref{eqn:cont}.

The continuity results are shown in fig. \ref{fig:prerollcont}. It
can be seen that \emph{LOpt} and \emph{AMP} achieves similar results.
This is expected as \emph{AMP} was tuned to achieve a high continuity.
The performance disparity between \emph{AMP} and \emph{AMP25} shows
that this simulation requires a greater amount of playout slowdown
at the decoder in order to reduce the occurrences of buffer underflows.
However, both \emph{LOpt} and \emph{AMP} require about a 100 times
less prebuffering compared to \emph{Norm} to provide similar continuity.
\emph{AMP25} too requires about 7 times less prebuffering than \emph{Norm}
but still requires about 50 times more prebuffering than \emph{LOpt}
and \emph{AMP}. This suggests that using some form of playout slowdown
would reduce the prebuffering requirements of a video application
significantly. 

We next measure the playout distortion of each scheme using \eqref{eqn:h}
($p_{n}=1/30$), i.e. playout distortion will be produced when the
playout interval drops below or goes above $1/30$. Note that while
\eqref{eqn:h} will only produce a non-zero value when the playout
deviates from the natural playout frame interval, playout interruptions
due to buffer underflows are not factored into the playout distortion.
Fig. \ref{fig:prerolldist} shows the playout distortion results.
\emph{Norm} does not have any playout distortion since it has a constant
30 fps playout rate, but suffers from playout discontinuity as discussed
earlier. \emph{LOpt} has a very similiar playout distortion characteristic
when compared to \emph{AMP25}. In contrast, \emph{AMP} for most cases
produces twice the amount of playout distortions compared to \emph{LOpt}
and \emph{AMP25}. This is mainly because \emph{AMP} requires a higher
slowdown factor to obtain a better video continuity and, as a consequence,
this results in a higher playout distortion.

\emph{LOpt}'s comparatively low playout distortion is due to the joint
adjustment of both the encoder frame generation rate and playout rate.
By increasing the encoder frame generation rate, the receiving frame
at the decoder increases. This provides a higher buffer occupancy
and reduces the need to slowdown the playout rate, thus reducing the
playout distortion. This comes at the expense of frame quality, because
increasing encoder frame generation rate will result in a higher amount
of compression. To examine \emph{LOpt}'s effect on frame quality,
we compare the PSNR of the encoded video before transmission. This
is done to eliminate any possible drops in PSNR due to transmission.
\emph{Norm}, \emph{AMP25} and \emph{AMP} have a constant encoding
rate of 30 fps, this means all produce an encoded video of the same
PSNR. Thus, we only compared \emph{LOpt} with \emph{Norm}. From fig.
\ref{fig:prerollpsnr}, it is shown that the drop in PSNR is about
0.6 dB for \emph{LOpt}. This is a reasonably small tradeoff in frame
quality given the improvements in playout distortion and continuity. 

\begin{figure}[tbh]
\begin{minipage}[t]{0.45\textwidth}%
\includegraphics[scale=0.45]{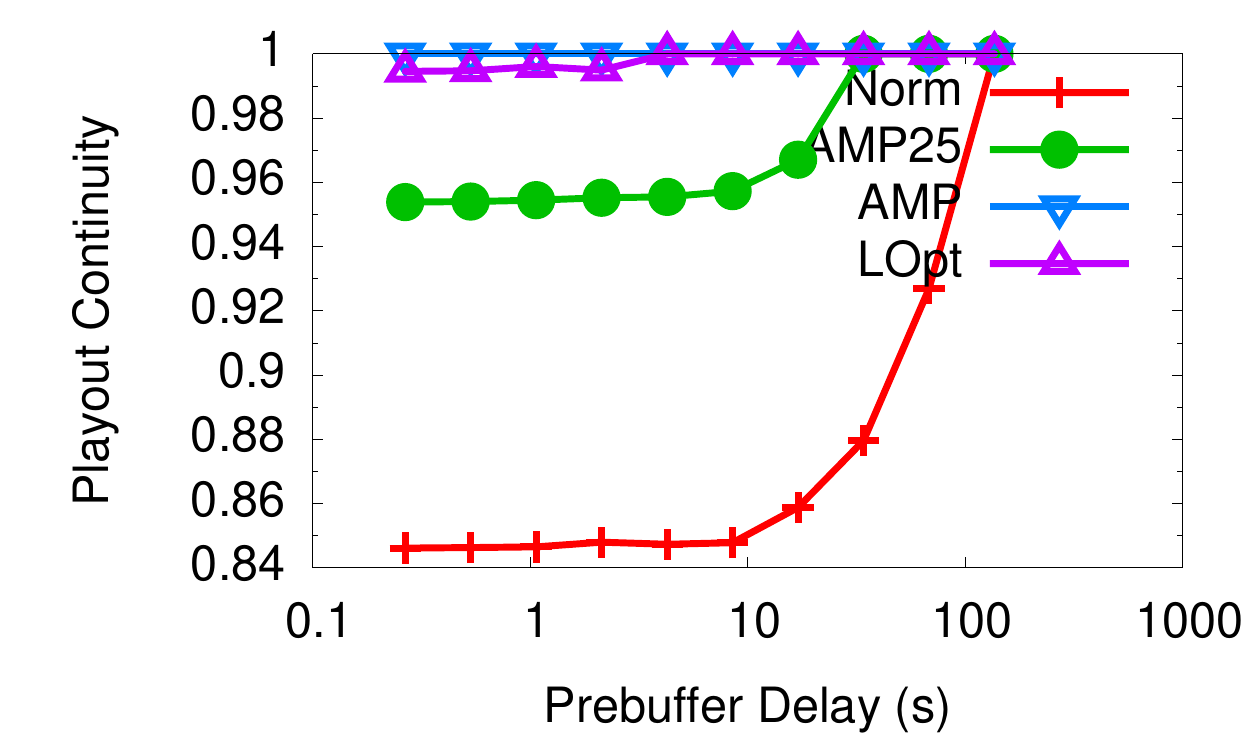}

\caption{Prebuffering delay vs continuity (football).}

\label{fig:prerollcont-football}%
\end{minipage}\hfill{}%
\begin{minipage}[t]{0.45\textwidth}%
\includegraphics[scale=0.45]{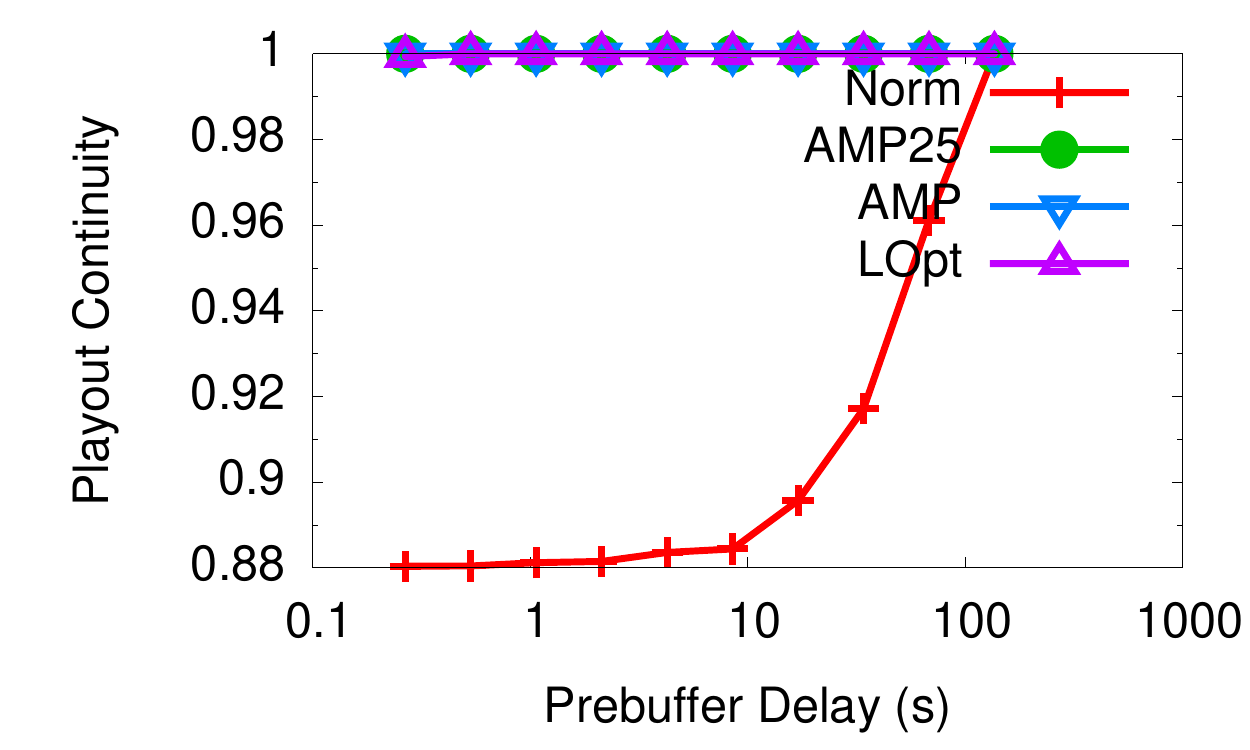}

\caption{Prebuffering delay vs continuity (akiyo). }
\label{fig:prerollcont-akiyo}%
\end{minipage}\hfill{}%
\begin{minipage}[t]{0.45\textwidth}%
\includegraphics[scale=0.45]{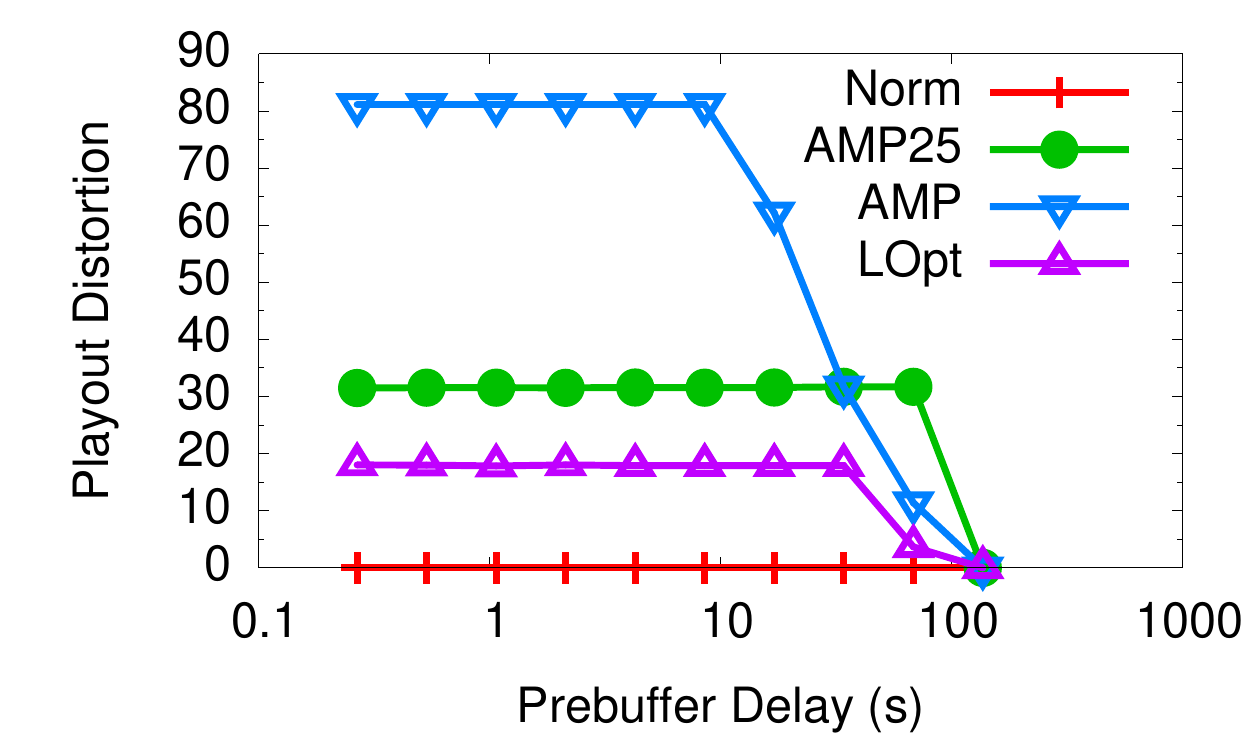}

\caption{Prebuffering delay vs playout distortion (football). }
\label{fig:prerolldist-football}%
\end{minipage}\hfill{}%
\begin{minipage}[t]{0.45\textwidth}%
\includegraphics[scale=0.45]{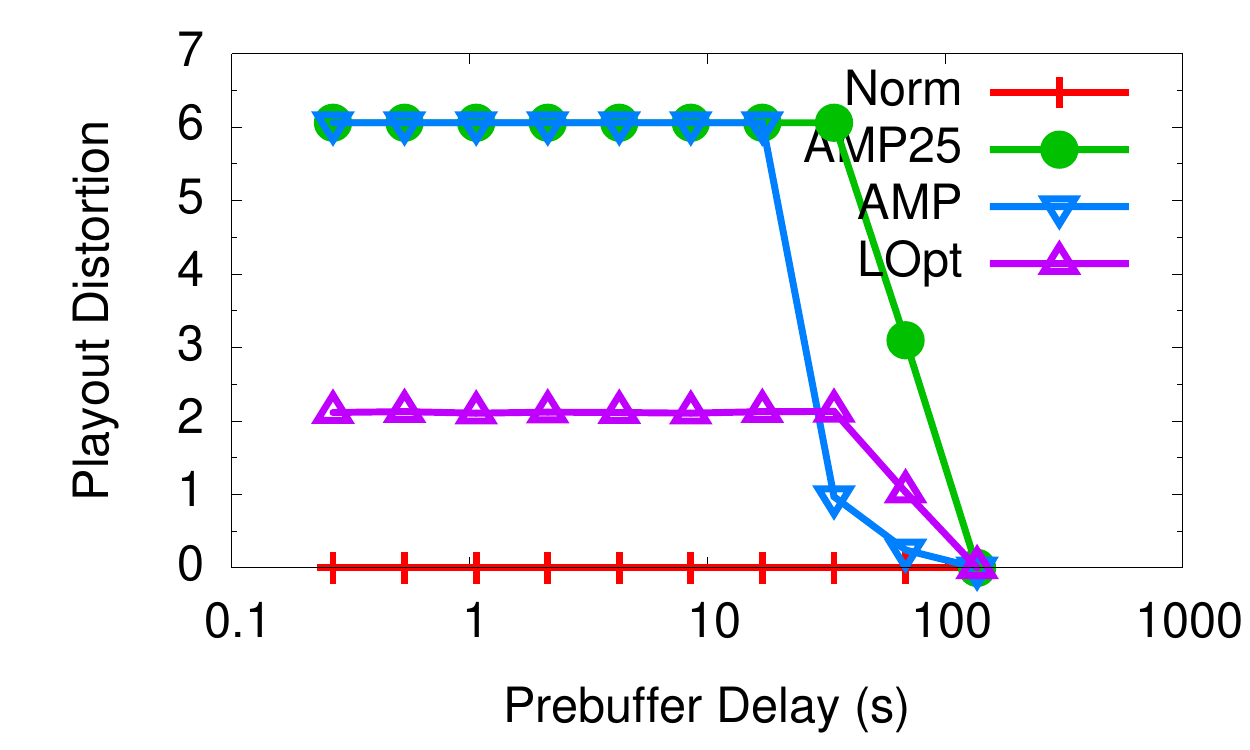}

\caption{Prebuffering delay vs playout distortion (akiyo). }
\label{fig:prerolldist-akiyo}%
\end{minipage}\hfill{}%
\begin{minipage}[t]{0.45\textwidth}%
\includegraphics[scale=0.45]{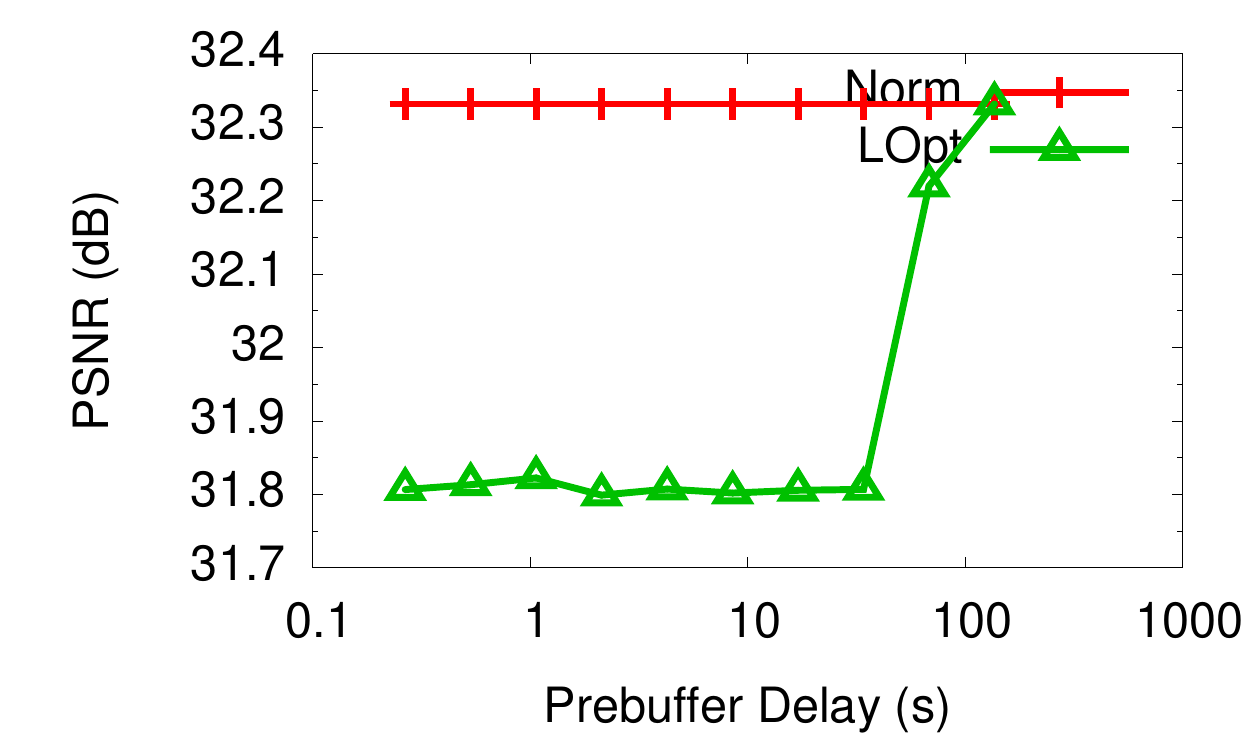}

\caption{PSNR loss for \emph{LOpt} (football).}
\label{fig:prerollpsnr-football}%
\end{minipage}\hfill{}%
\begin{minipage}[t]{0.45\textwidth}%
\includegraphics[scale=0.45]{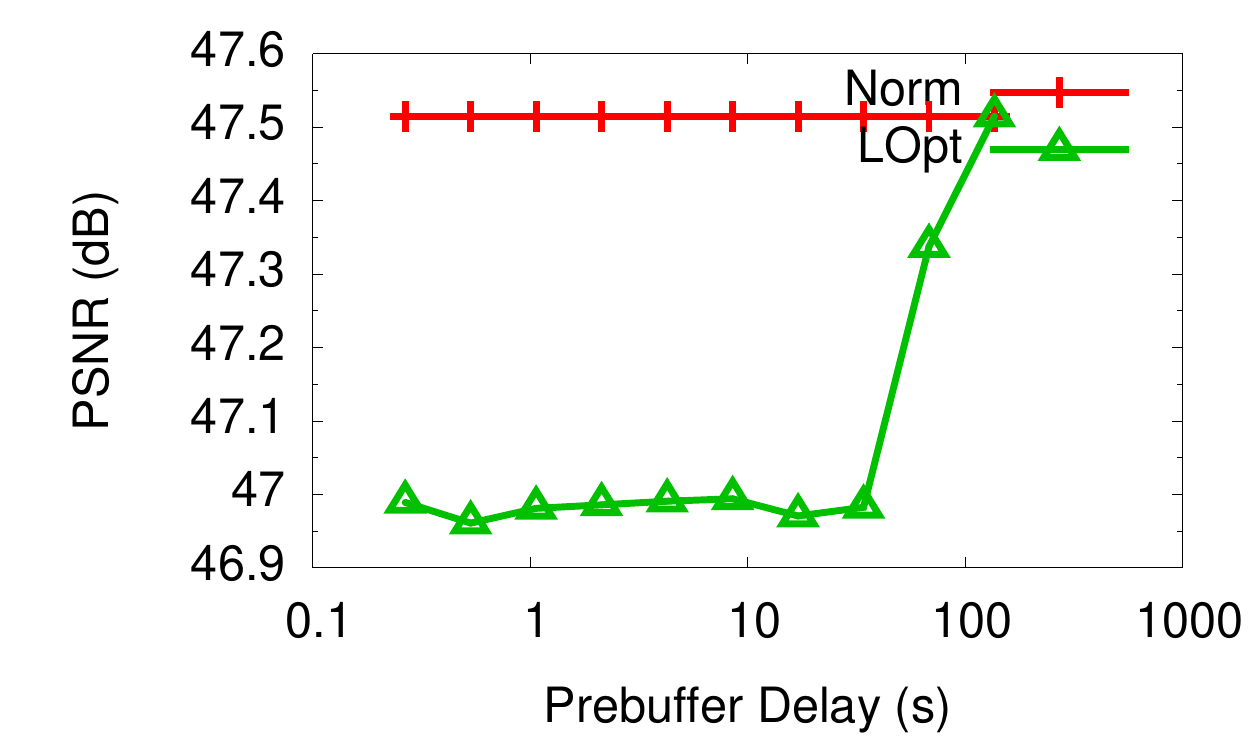}

\caption{PSNR loss for \emph{LOpt} (akiyo).}
\label{fig:prerollpsnr-akiyo}%
\end{minipage}
\end{figure}

It can be also seen from the graphs that the performance of \emph{LOpt}
and \emph{LOptM} are very similiar. This suggests that the bound in
\eqref{eqn:b3} is not too conservative. We also tested the schemes
on football, see figs. \ref{fig:prerollcont-football}, \ref{fig:prerolldist-football},
\ref{fig:prerollpsnr-football} and akiyo, see figs. \ref{fig:prerollcont-akiyo},
\ref{fig:prerolldist-akiyo}, \ref{fig:prerollpsnr-akiyo}. football
and akiyo were chosen because they have the highest and lowest motion
content respectively in the four sequences used. The results show
a similiar pattern to the concatenated sequence.

We now examine the complexity of each scheme.\emph{ Norm} is the least
complex scheme while the \emph{AMP25} is marginally more complex than
\emph{Norm}. This is because \emph{AMP25} involves some simple logic
at the decoder to slowdown the playout once a certain buffer level
is reached. \emph{LOpt }is more complex than \emph{AMP25} as it involves
more calculations at both the encoder and decoder end. However, it
is not significantly more complex than \emph{AMP25} because the optimization
policies are concave, so the solution search is very efficient. For
example, in our implementation we solve for $f(t)$ and $p(t)$ by
using the first derivatives of the encoder policy \eqref{eqn:encobj}
and decoder policy \eqref{eqn:decobj} respectively. \emph{AMP} is
the most complex in our implementation as it requires several runs
to determine the most optimal slowdown factor. In summary, the \emph{LOpt}
runs in $O(n)$, where $n$ is the total number of video packets.
While \emph{AMP} runs in $O(n\times s)$, where $s$ is the number
of slowdown factors to consider. However, it should be noted that
complexity-wise, \emph{AMP} is not representative of the complexity
of AMP schemes in practice. Its main purpose in our simulations is
to act as the upper bound in the performance of AMP schemes.

\subsection{Delay Constrained Lyapunov Optimization Results\label{sub:Delay-Constrained-Exp}}

\begin{figure}[tbh]
\begin{minipage}[t]{0.45\textwidth}%
\includegraphics[scale=0.5]{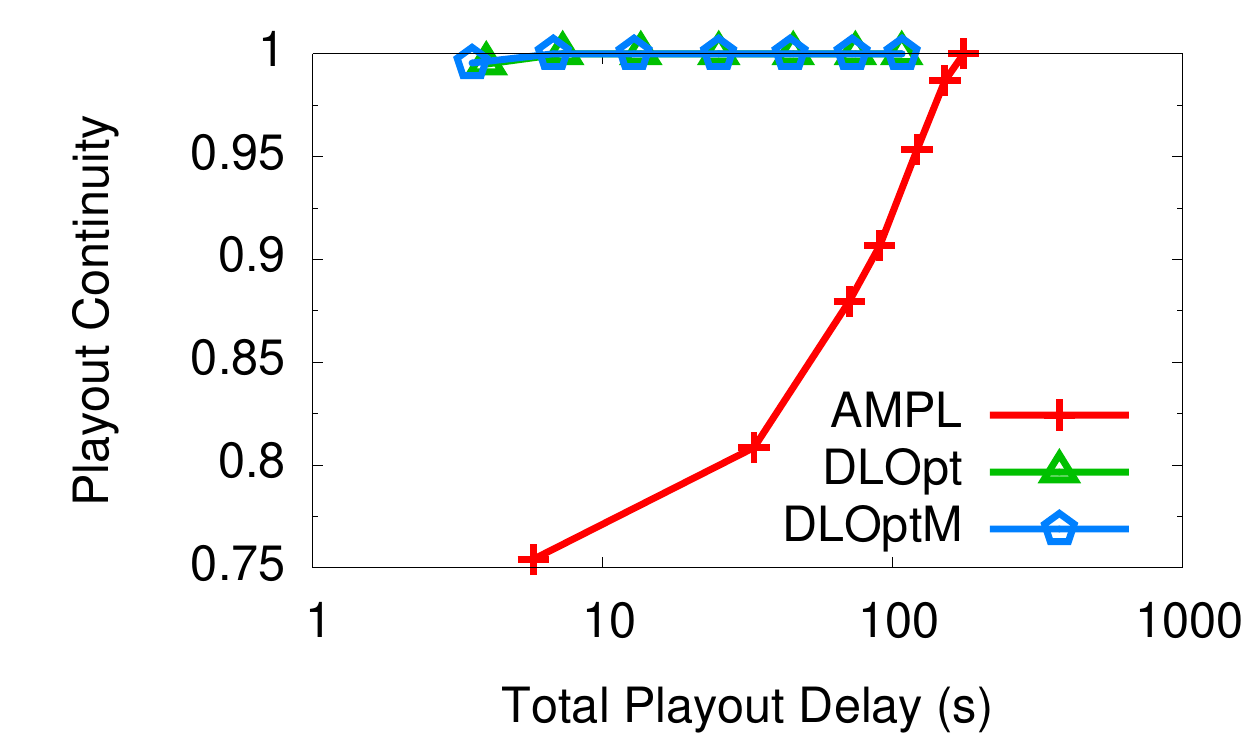}

\caption{Playout delay vs continuity.}
\label{fig:livecont}%
\end{minipage}\hfill{}%
\begin{minipage}[t]{0.45\textwidth}%
\includegraphics[scale=0.5]{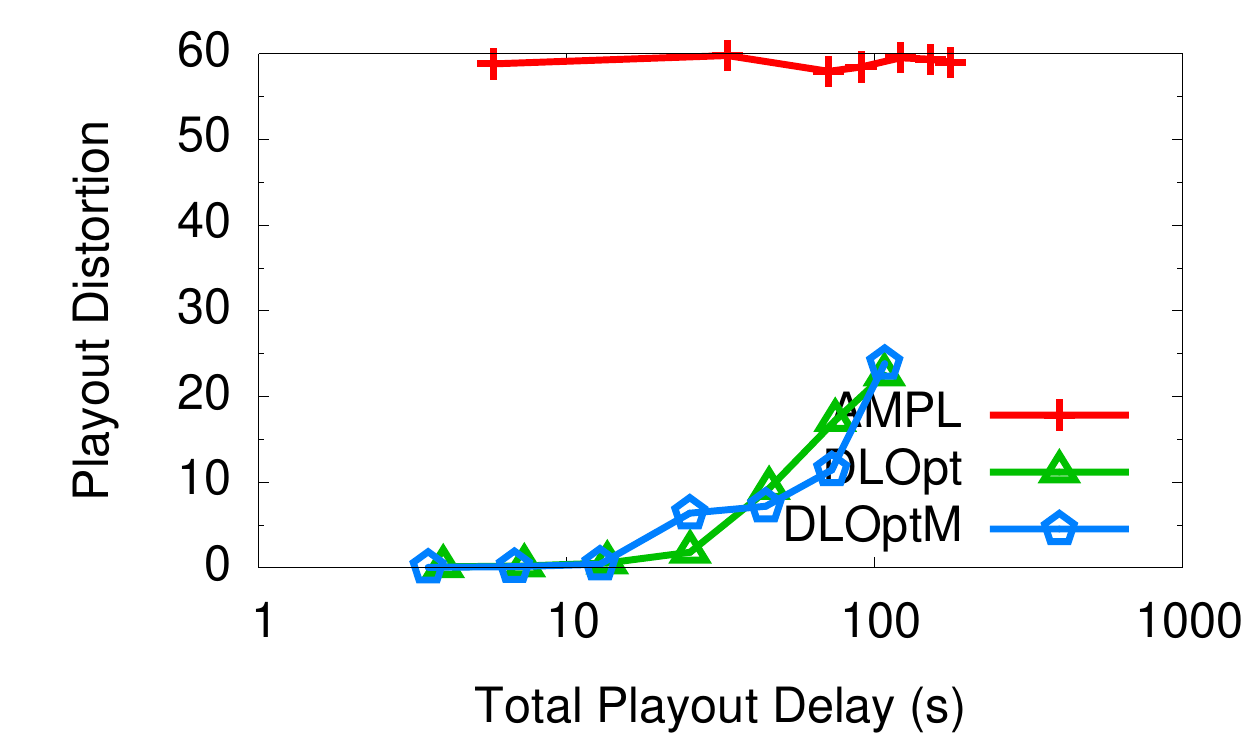}

\caption{Playout delay vs playout distortion.}
\label{fig:livedist}%
\end{minipage}\hfill{}%
\begin{minipage}[t]{0.45\textwidth}%
\includegraphics[scale=0.5]{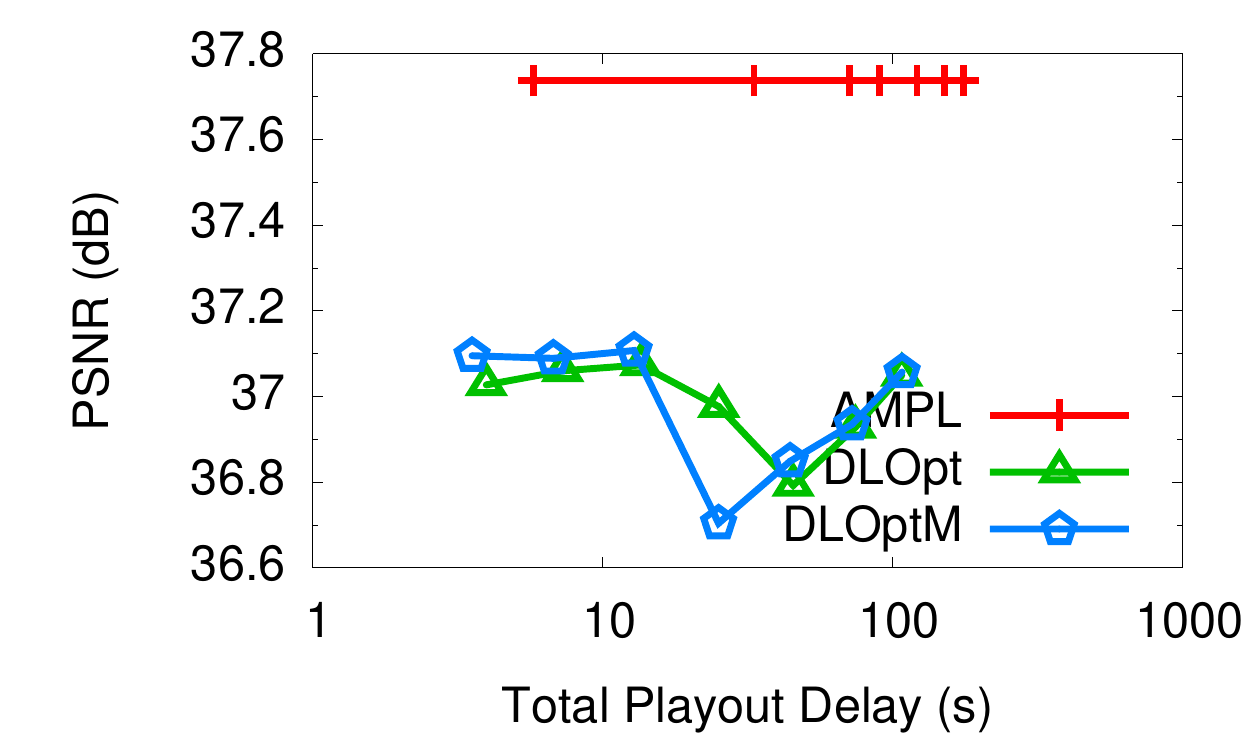}

\caption{PSNR loss for \emph{DLOpt}}
\label{fig:livepsnr}%
\end{minipage}
\end{figure}

To evaluate the performance of our delay constrained Lyapunov optimization
framework, which we call \emph{DLOpt} here. As in the previous section,
we also set up out framework with a less conservative bound $\frac{b(t)\, r_{max}\, p_{max}}{T-t}$,
labelled as \emph{DLOptM}. We compare our scheme with AMP-Live \cite{ref:ampkalman},
labelled as \emph{AMPL}. AMP-Live maintains the buffer level by slowing
down or speeding up the playout based on a predetermined scale factor.
The scale factor of \emph{AMPL} is set to 40\% in our experiments,
this value was found to provide the best overall performance for \emph{AMPL}.
The playout delay of each scheme is measured as: $\text{Total playout delay}=\sum_{\tau=0}^{T}(p(\tau)-p_{n})$.

Recall that $p_{n}$ represents the natural playout frame interval
of the sequence. So every playout slowdown will cause $p(t)$ to be
larger than $p_{n}$, thus accumulating playout delay. To reduce the
total playout delay, the scheme needs to find the proper moment to
increase the playout rate.

We compared the schemes by varying the delay constraints from 4.27
seconds to 273.07 seconds with the constraint doubled at each run.
At the end of each run, we plot the total playout delay and the performance
metric of each scheme, the playout delay in the results are on a log
base 10 scale. The performance metric we examined are playout continuity,
playout distortion and PSNR.

We first look at the continuity results with respect to the total
playout delay (fig. \ref{fig:livecont}). It can be seen that \emph{DLOpt}
achieves maximum continuity regardless of the delay constraint, while
\emph{AMPL} requires more than 50 times the amount of total playout
delay compared to \emph{DLOpt} to reach maximum continuity. \emph{AMPL}
manages the total playout delay constraint by maintaining a certain
buffer level \cite{ref:ampkalman}, thus a tighter delay constraint
will result in a lower buffer level which has higher probabilities
of buffer underflows. On the other hand, \emph{DLOpt} makes a trade-off
between the encoder frame generation rate and playout rate to satisfy
the continuity goal and delay constraint.

We next look at the distortion results in fig. \ref{fig:livedist}.
Again, \emph{DLOpt} achieves a much lower playout distortion compared
to \emph{AMPL}. Note that the lower the total playout delay constraints
the lower the playout distortion. This is because a low playout constraint
causes \emph{DLOpt} to adjust the playout rate a lot less, thus causing
a low playout distortion. \emph{AMPL} has an almost constant playout
distortion. This is because, as mentioned before, \emph{AMPL} only
tries to maintain a certain buffer level and does not constrain its
playout rate adjustment in anyway.

Lastly, we examine the PSNR trade-offs made, see fig. \ref{fig:livepsnr}.
As before, we compared the PSNR of the encoded video of both schemes
prior to transmission. It can be seen that \emph{DLOpt} sacrifices
about a maximum of 1 dB of PSNR, this PSNR drop is higher than \emph{LOpt}
in the previous section. The main cause of this is the additional
delay constraint imposed using the virtual buffer \eqref{eqn:Xbuff}.
This results a lower PSNR due to the need to satisfy the delay constraint,
however, this allows the large gains in continuity and distortion.

Again the performances of \emph{DLOpt} and \emph{DLOptM} are mostly
similiar. This reinforces the possibility that the bound in \eqref{eqn:b3}
is not too conservative..

\section{Conclusions\label{sec:Conclusions}}

In this paper, we proposed a the frame rate optimization framework.
What we achieved with this framework is:
\begin{itemize}
\item Performed frame rate control by a joint adjustment of the encoder
frame generation interval and the playout frame interval.
\item Modelled the frame rate control problem using the encoder frame generation
interval, the playout frame interval and the discontinuity penalty
virtual buffer. The model is created in such a way that stabilizing
the discontinuity penalty virtual buffer allows the video to maintain
continuity. We then used Lyapunov optimization on this model to systematically
derive the optimization policies. We showed that these policies can
be decoupled into separate encoder and decoder optimization policies
with feedback between the two. We also showed through experiments
that the proposed discontinuity penalty based on the virtual buffer
is correlated to the video continuity. Finally, simulation results
demonstrates the effectiveness of the discontinuity penalty virtual
buffer approach.
\item A delay constraint imposed on the accumulated delay from playout slowdowns.
We introduced the delay constraint into the framework using the delay
accumulator virtual buffer. We showed, using Lyapunov optimization
analysis, that by stabilizing the delay accumulator virtual buffer,
the delay constraint would be satisfied. Simulation results showed
a superior playout continuity and playout distortion performance with
a reasonable tradeoff in PSNR. 
\item An analysis of the impact of delayed feedback from receiver to sender.
We derived two different analyses, the first analysis showed very
little impact on the optimization polices. The alternate analysis
showed that the decoder needed to use an outdated buffer state. Simulation
results demonstrated that using the first analysis results in a better
performance.
\end{itemize}
\bibliographystyle{IEEEtran}
\bibliography{bufcom}

\pagebreak{}

\appendix

\subsection{Derivation of Discontinuity Penalty Lyapunov Drift Bound\label{sec:apx-drift1-froimp}}

We square the buffer dynamics equation \eqref{eqn:vb2}, divide it
by two and rearrange its terms to get:

\begin{align}
U(t+1) & \leq[U(t)-p(t)-\beta(t)]^{+}+F(f(t))\notag\\
U(t+1)^{2} & \leq U(t)^{2}+\left(p(t)+\beta(t)\right)^{2}+F(f(t))^{2}\notag\\
 & \quad-2U(t)\left[p(t)+\beta(t)-F(f(t))\right]\end{align}

Dividing the above equation by two and rearranging, we get:

\begin{equation}
\frac{U(t+1)^{2}}{2}-\frac{U(t)^{2}}{2}\leq\frac{\left(p(t)+\beta(t)\right)^{2}+F(f(t))^{2}}{2}-U(t)\left[p(t)+\beta(t)-F(f(t))\right]\label{eq:apx-drift1-0}\end{equation}

Using the definition of the Lyapunov function \eqref{eqn:lyap} in
the above equation yields:

\begin{align}
L(U(t+1)-L(U(t)) & \leq\frac{\left(p(t)+\beta(t)\right)^{2}+F(f(t))^{2}}{2}\notag\label{eq:apx-drift1-1}\\
 & \quad-U(t)\left[p(t)+\beta(t)-F(f(t))\right]\end{align}

Recall that $\beta(t)$ is defined as:

\begin{equation}
\beta(t)=\frac{b(t)\, r_{min}\, p_{min}}{T-t}\end{equation}

where $b(t)$ is the amount of buffered video frames in the buffer
at time $t$ and $T$ is the length of the whole sequence in time
slots. Since a time slot corresponds to a frame in our framework,
it can be seen that $b(t)\leq T$. This is because the maximum amount
of video content that can exist in the buffer is the whole sequence.
This then implies that:

\begin{equation}
\beta(t)\leq\frac{T\, r_{min}\, p_{min}}{T-t}\label{eq:apx-beta-bnd}\end{equation}

Using the above bound in \eqref{eq:apx-drift1-1}, we obtain:

\begin{align}
L(U(t+1)-L(U(t)) & \leq\frac{1}{2}\left[\left(p(t)+\frac{T\, r_{min}\, p_{min}}{T-t}\right)^{2}+F(f(t))^{2}\right]\notag\\
 & \quad-U(t)\left[p(t)+\beta(t)-F(f(t))\right]\end{align}

Using $r(t)=F(f(t))$ and $r(t)\leq r_{max}$, we get $F(f(t))\leq r_{max}$.
Since $p(t)\leq p_{max}$, using these two bounds into the above equation
yields:

\begin{align}
L(U(t+1)-L(U(t)) & \leq\frac{1}{2}\left[\left(p_{max}+\frac{T\, r_{min}\, p_{min}}{T-t}\right)^{2}+r_{max}^{2}\right]\notag\\
 & \quad-U(t)\left[p(t)+\beta(t)-F(f(t))\right]\notag\\
 & =B-U(t)\left[p(t)+\beta(t)-F(f(t))\right]\label{eq:apx-drift1-2}\end{align}

where:

\begin{equation}
B=\frac{1}{2}\bigg(r_{max}^{2}+\bigg(p_{max}+\frac{T\, r_{min}\, p_{min}}{T-t}\bigg)^{2}\bigg)\end{equation}

If we take conditional expectations of \eqref{eq:apx-drift1-2} with
respect to $U(t)$ and use the definition of the one-step conditional
Lyapunov drift \eqref{eqn:1stepdrift}, we get:

\begin{equation}
\Delta(U(t))\leq B-U(t)\mathbb{E}\{\beta(t)+p(t)-F(f(t))|U(t)\}\end{equation}

\subsection{Derivation of Delay Constrained Lyapunov Drift Bound\label{sec:apx-Xdrift1-froimp}}

We first define a Lyapunov function $L(X(t))$ based on $X(t)$ as:

\begin{equation}
L(X(t))\triangleq\frac{X^{2}(t)}{2}\label{eqn:Xonlylyap}\end{equation}

Recall for the discontinuity penalty $U(t)$, its Lyapunov function
is:

\begin{equation}
L(U(t))\triangleq\frac{U^{2}(t)}{2}\end{equation}

Then, it can be seen that \eqref{eqn:Xlyap} can be defined as:

\begin{align}
L(U(t),X(t)) & =\frac{U^{2}(t)+X^{2}(t)}{2}\notag\\
 & =\frac{U^{2}(t)}{2}+\frac{X^{2}(t)}{2}\notag\\
 & =L(U(t))+L(X(t))\label{eq:apx-comblyap}\end{align}

Now, we square the buffer dynamics equation \eqref{eqn:Xbuff} to
obtain:

\begin{align}
X(t+1) & \leq[X(t)-p_{n}-t_{d}]^{+}+p(t)\notag\\
X(t+1)^{2} & \leq X(t)^{2}+p(t)^{2}+\left(p_{n}+t_{d}\right)^{2}-2X(t)\left[p_{n}+t_{d}-p(t)\right]\end{align}

Dividing the above equation by two and rearranging yields:

\begin{equation}
\frac{X(t+1)^{2}}{2}-\frac{X(t)^{2}}{2}\leq\frac{p(t)^{2}+\left(p_{n}+t_{d}\right)^{2}}{2}-X(t)\left[p_{n}+t_{d}-p(t)\right]\end{equation}

Using $p(t)\leq p_{max}$ and \eqref{eqn:Xonlylyap} in the above
equation yields:

\begin{align}
L(X(t+1))-L(X(t)) & \leq\frac{1}{2}\left(p(t)^{2}+\left(p_{n}+t_{d}\right)^{2}\right)-X(t)\left[p_{n}+t_{d}-p(t)\right]\notag\\
 & =C-X(t)\left[p_{n}+t_{d}-p(t)\right]\label{eq:apx-x-1}\end{align}

where:

\begin{equation}
C=\frac{1}{2}\bigg(p_{max}^{2}+(p_{n}+t_{d})^{2}\bigg)\end{equation}

By adding \eqref{eq:apx-x-1} and \eqref{eq:apx-drift1-2} together,
we get:

\begin{align}
L(X(t+1)) & -L(X(t))+L(U(t+1)-L(U(t))\leq & \notag\\
 & C-X(t)\left[p_{n}+t_{d}-p(t)\right]+B-U(t)\left[p(t)+\beta(t)-F(f(t))\right]\end{align}

Using \eqref{eq:apx-comblyap} and rearranging the above yields:

\begin{align}
L(U(t+1),X(t+1)) & -L(U(t),X(t))\leq B+C & \notag\\
 & \quad-U(t)\left[p(t)+\beta(t)-F(f(t))\right] & \notag\\
 & \quad-X(t)\left[p_{n}+t_{d}-p(t)\right]\end{align}

Taking conditional expectations of the above with respects to $U(t)$
and $X(t)$, and using \eqref{eqn:Xonestep} yields:

\begin{align}
\Delta(U(t),X(t))\leq & \; B+C\notag\\
 & -U(t)\mathbb{E}\{p(t)-F(f(t))|U(t),X(t)\}\notag\\
 & -X(t)\mathbb{E}\{p_{n}+t_{d}-p(t)|U(t),X(t)\}\end{align}

\subsection{Derivation of Network Delayed Lyapunov Drift Bound\label{sec:apx-deldrift1-froimp}}

Note that, in this section, we use the same shortened notations defined
in Section \ref{sec:Network-Delay-Impact}, additionally we also define
$U_{d_{f}}=U(t+d_{f}+1)$. This means that \eqref{eqn:currfut} can
be rewritten as:

\begin{equation}
U_{d_{f}}\leq\left[U-\gamma_{d_{f}}\right]^{+}+F_{d_{f}}\end{equation}

By squaring the above equation and dividing it by two, we obtain:

\begin{equation}
\frac{U_{d_{f}}^{2}}{2}\leq\frac{U^{2}}{2}+\frac{\gamma_{d_{f}}^{2}+F_{d_{f}}^{2}}{2}-U\left[\gamma_{d_{f}}-F_{d_{f}}\right]\end{equation}

Using the definition of the Lyapunov function \eqref{eqn:lyap} on
the above equation and rearranging:

\begin{equation}
L\left(U_{d_{f}}\right)-L(U)\leq\frac{\gamma_{d_{f}}^{2}+F_{d_{f}}^{2}}{2}-U\left[\gamma_{d_{f}}-F_{d_{f}}\right]\end{equation}

Using $\gamma(t)=p(t)+\beta(t)$, $F(f(t))\leq r_{max}$%
\footnote{See Section \ref{sec:apx-drift1-froimp}.%
}, $p(t)\leq p_{max}$ and \eqref{eq:apx-beta-bnd} on the above equation
yields:

\begin{align}
L\left(U_{d_{f}}\right)-L(U) & \leq\frac{d_{f}}{2}\bigg(r_{max}^{2}+\bigg(p_{max}+\frac{T\, r_{min}\, p_{min}}{T-t}\bigg)^{2}\bigg)\notag\\
 & =B'-U\left[\gamma_{d_{f}}-F_{d_{f}}\right]\label{eq:apx-deldrift-1}\end{align}

where:

\begin{equation}
B'=\frac{d_{f}}{2}\bigg(r_{max}^{2}+\bigg(p_{max}+\frac{T\, r_{min}\, p_{min}}{T-t}\bigg)^{2}\bigg)\end{equation}

Taking conditional expectations of \eqref{eq:apx-deldrift-1} with
respects to $U$ and using \eqref{eqn:dfstepdrift} yields:

\begin{equation}
\Delta(U)\leq B'-\mathbb{E}\left\{ U\gamma_{d_{f}}\big|U\right\} +\mathbb{E}\left\{ UF_{d_{f}}\big|U\right\} \end{equation}

\subsection{Discontinuity Penalty Optimization Stability and Performance Bounds\label{sub:Discontinuity-Penalty-bnd}}

In this section, we prove that the derived encoder and decoder policies,
\eqref{eqn:encobj} and \eqref{eqn:decobj} respectively, stabilize
the discontinuity penalty virtual buffer. This is given by equation
\eqref{eqn:dpthm_ubnd} in Theorem \ref{thm:discontinuity-penalty-policies}.
We also show the performance bound of the Lyapunov optimization. 
\begin{thm}
\label{thm:discontinuity-penalty-policies}Let the long term receiving
frame interval\textup{ $\bar{r}=\lim_{t\to\infty}\frac{1}{t}\sum_{\tau=0}^{t-1}\mathbb{E}\left\{ r(t)\right\} $}.
If the network delay variation $r(t)$ is i.i.d%
\footnote{With i.i.d processes, the steady state is exactly achieve every timeslot.
This allows us to use Lyapunov drift analysis on a per time slot basis.
However, Neely has shown that i.i.d processes provides all of the
intuition needed to treat general processes. For more details on this,
see chapter 4 of \cite{ref:neelythesis}.%
} over the timeslots, the receiving frame interval is lower bounded
by the encoder frame generation interval as \textup{$r(t)\geq f(t)$},
the frame quality function is bounded as $g(f(t))\leq G_{max}$ and
the playout distortion function is bounded as $h(p(t))\geq H_{min}$.
Then implementing the optimization policies \eqref{eqn:encobj} and
\eqref{eqn:decobj} in each timeslot stabilizes the discontinuity
penalty $U(t)$ (using the stability definition \textup{$\mathbb{E}\{U\}\triangleq\limsup_{t\to\infty}\frac{1}{t}\sum_{\tau=0}^{t-1}\mathbb{E}\{U(\tau)\}<\infty$})
and satisfies the following performance bounds:

\begin{equation}
\limsup_{M\to\infty}\frac{1}{M}\sum_{\tau=0}^{M-1}\mathbb{E}\{U(\tau)\}\leq\frac{B+V(G_{max}-H_{min})}{r_{max}}\label{eqn:dpthm_ubnd}\end{equation}

\begin{equation}
\liminf_{M\to\infty}g(\bar{f})\geq(g^{*}-h^{*})+H_{min}-\frac{B}{V}\label{eqn:dpthm_gbnd}\end{equation}

\begin{equation}
\limsup_{M\to\infty}h(\bar{p})\leq\frac{B}{V}-(g^{*}-h^{*})-G_{max}\label{eqn:dpthm_hbnd}\end{equation}

where $V$ is some positive constant (i.e. $V>0$), $g^{*}$ and $h^{*}$
are the specific values of $g(.)$ and $h(.)$ respectively that maximizes
the objective \eqref{eqn:lyapoptobj} subjected to the constraints
\eqref{eqn:lyapoptobj1}, \eqref{eqn:lyapoptobj2} and \eqref{eqn:lyapoptobj3},
B is defined as in \eqref{eqn:driftB} and:

\begin{equation}
\overline{f}=\frac{1}{M}\sum_{\tau=0}^{M-1}\mathbb{E}\{f(\tau)\}\end{equation}

\begin{equation}
\overline{p}=\frac{1}{M}\sum_{\tau=0}^{M-1}\mathbb{E}\{p(\tau)\}\end{equation}
\end{thm}
\begin{IEEEproof}
We first introduce $\Lambda$ as a set of receiving frame intervals
that stabilizes the discontinuity penalty $U(t)$. $\Lambda$ is bounded
by $r_{max}$. Using $\Lambda$ assumes a complete knowledge of future,
but this is only required for the analysis on the performance bounds.
With $\Lambda$, the optimization problem in Section \ref{sub:Virtual-Buffer}
can be restated as:

\begin{align}
\mathrm{\textrm{Maximize:}} & \quad g(f(t))-h(p(t))\label{eqn:fro-imp-proofobj1}\\
\textrm{Subject to:} & \quad F(f(t))=r(t)\label{eqn:fro-imp-proofcon1}\\
 & \quad\bar{r}=\lim_{t\to\infty}\frac{1}{t}\sum_{\tau=0}^{t-1}\mathbb{E}\left\{ r(t)\right\} \\
 & \quad\bar{r}\in\Lambda\\
 & \quad f_{min}\leq f(t)\leq f_{max}\\
 & \quad p_{min}\leq p(t)\leq p_{max}\end{align}

From the above optimization problem, $\Lambda$ can be intuitively
seen as a range of values that $r(t)$ can take that allows the existence
of at least one stationary randomized policy that can stabilize $U(t)$.
Constraint \eqref{eqn:fro-imp-proofcon1} is due to $r(t)=F(f(t))$.
Let $f^{*}$ and $p^{*}$ form the solution that maximizes the objective
\eqref{eqn:fro-imp-proofobj1}. This means that an optimal stable
policy would ensure that the following is met:

\begin{equation}
F(f^{*})=\bar{r}\leq p^{*}\label{eqn:fro-imp_proofprop0}\end{equation}

$F(f^{*})=\bar{r}$ comes from constraint \eqref{eqn:fro-imp-proofcon1},
while the inequality $\bar{r}\leq p^{*}$ is the result of lemma 7
in \cite{ref:neelyvb}. Suppose now an $\epsilon$-optimal stable
policy causes the long term receiving frame interval $\bar{r}$ to
become $\bar{r}_{\epsilon}$ such that:

\begin{align}
F(f_{\epsilon}^{*})=\bar{r}_{\epsilon}\leq p^{*}-\epsilon & \notag\label{eqn:fro-imp_proofprop1}\\
\bar{r}_{\epsilon}\in\Lambda_{\epsilon} & \notag\\
\bar{r}_{\epsilon}+\epsilon\in\Lambda\end{align}

where $\epsilon>0$ is a positive constant and the receiving frame
intervals set $\Lambda_{\epsilon}\subset\Lambda$. $\Lambda_{\epsilon}$
can be seen as $\Lambda$ that is reduced by $\epsilon$. Then, the
$\epsilon$-optimal policy can be seen as optimizing the following
problem:

\begin{align}
\mathrm{\textrm{Maximize:}} & \quad g(f(t))-h(p(t))\notag\\
\textrm{Subject to:} & \quad F(f(t))=r(t)\notag\\
 & \quad r(t)\in\Lambda_{\epsilon}\notag\\
 & \quad f_{min}\leq f(t)\leq f_{max}\notag\\
 & \quad p_{min}\leq p(t)\leq p_{max}\label{eq:apx-froimp-thm-eppol}\end{align}

Note that the long term encoder frame generation interval $f_{\epsilon}^{*}$
forms part of the solution that maximizes the above problem. Equations\eqref{eqn:fro-imp_proofprop0}
and \eqref{eqn:fro-imp_proofprop1} together implies that:

\begin{equation}
\bar{r}_{\epsilon}\leq\bar{r}\end{equation}

This is because the $\epsilon$-optimal policy maximizes the same
objective as \eqref{eqn:fro-imp-proofobj1} but with its long term
receiving frame interval $\bar{r}_{\epsilon}$ limited by $\epsilon$.
Using $r(t)=F(f(t))$ would mean the above equation becomes:

\begin{equation}
F(f_{\epsilon}^{*})\leq F(f^{*})\end{equation}

Using \eqref{eqn:Fdefine} and assuming there exist the long term
frame interval scaling factors $e_{\epsilon}$ and $e^{*}$, such
that $F(f_{\epsilon}^{*})=e_{\epsilon}f_{\epsilon}^{*}=\bar{r}_{\epsilon}$
and $F(f^{*})=e^{*}f^{*}=\bar{r}$. Then, applying these properties
to the above equation yields:

\begin{equation}
e_{\epsilon}f_{\epsilon}^{*}\leq e^{*}f^{*}\end{equation}

Since we made the assumption that $r(t)\geq f(t)$, which can be rewritten
as $e(t)f(t)\geq f(t)$. This implies that $e_{\epsilon}\geq1$ and
$e^{*}\geq1$, which means that :

\begin{equation}
f_{\epsilon}^{*}\leq f^{*}\label{eqn:fro-imp-proofutildec}\end{equation}

This is because $f_{\epsilon}^{*}$ is chosen by the $\epsilon$-optimal
policy to maximize the objective in \eqref{eq:apx-froimp-thm-eppol}.
Since the frame quality function $g(f(t))$ in the objective is an
increasing function of $f(t)$, this implies that the $\epsilon$-optimal
policy will choose $f_{\epsilon}^{*}$ as large as possible. However,
as $f_{\epsilon}^{*}$ is upper bounded by $\bar{r}_{\epsilon}$,
which is in turn limited by $\epsilon$, this means that $f_{\epsilon}^{*}$
will be at most as large as $f^{*}$. Furthermore, $f_{\epsilon}^{*}\to f^{*}$
as $\epsilon\to0$. This is because \cite{ref:neelyvb}:

\begin{equation}
f^{*}\geq\left(1-\frac{\epsilon}{r_{max}}\right)f^{*}+\frac{\epsilon}{r_{max}}f_{\epsilon}^{*}\geq f_{\epsilon}^{*}\label{eqn:fro-imp_proofprop4}\end{equation}

The middle term of \eqref{eqn:fro-imp_proofprop4} can be seen as
an example of a mixed policy that picks $f^{*}$ with $1-\frac{\epsilon}{r_{max}}$
probability and $f_{\epsilon}^{*}$ with $\frac{\epsilon}{r_{max}}$
probability. 

If the long term receiving frame interval is $\bar{r}_{\epsilon}$,
then there exists a stationary randomized policy that stabilizes $U(t)$
by setting the long term playout interval to $p^{*}$ \cite{ref:neelythesis}.
That is with $\bar{r}_{\epsilon}$ as the long term average arrival
rate into the discontinuity penalty virtual buffer $U(t)$. If the
policy chooses the playout interval $p(t)$ over time such that the
long term receiving frame interval is $p^{*}$, then $U(t)$ can stabilize
since $p^{*}>\bar{r}_{\epsilon}$. Thus, it will not grow infinitely
large over time. Under such a policy, the Lyapunov one step drift
bound \eqref{eqn:lyapobj} would be calculated using \eqref{eqn:fro-imp_proofprop1}
as:

\begin{align}
\Delta(U(t)) & -V\mathbb{E}\{g(f(t))-h(p(t))|U(t)\} & \notag\\
\leq B & -\mathbb{E}\{U(t)\beta(t)|U(t)\} & \notag\\
 & -Vg(f_{\epsilon}^{*})+U(t)\bar{r}_{\epsilon} & \notag\\
 & -U(t)p^{*}+Vh(p^{*}) & \notag\\
\leq B & -\mathbb{E}\{U(t)\beta(t)|U(t)\} & \notag\\
 & -Vg(f_{\epsilon}^{*})+U(t)(p^{*}-\epsilon) & \notag\\
 & -U(t)p^{*}+Vh(p^{*})\end{align}

By rearranging the terms:

\begin{align}
\Delta(U(t)) & -V\mathbb{E}\{g(f(t))-h(p(t))|U(t)\}\notag\label{eqn:dpthm-lyapobj2}\\
\leq B & -V\left(g(f_{\epsilon}^{*})-h(p^{*})\right)-\epsilon U(t)\\
\leq B & -V\left(g(f_{\epsilon}^{*})-h^{*}\right)-\epsilon U(t)\end{align}

By taking expectations, summing over the timeslots $\tau\in[0\:..\: M-1]$
and using the non-negativity of $L(U(t))$ to drop the term $\mathbb{E}\{L(U(M-1))\}$,
we get:\begin{align}
-\mathbb{E}\{L(U(0))\}-V\sum_{\tau=0}^{M-1}\mathbb{E}\{g(f(\tau))-h(p(\tau))\}\notag\label{eqn:dpthm-bnd1}\\
\leq MB-\epsilon\sum_{\tau=0}^{M-1}\mathbb{E}\{U(\tau)\}-VM(g(f_{\epsilon}^{*})-h^{*})\end{align}

To prove the discontinuity penalty bound \eqref{eqn:dpthm_ubnd},
we divide \eqref{eqn:dpthm-bnd1} by $M\epsilon$ and rearrange its
terms to obtain:

\begin{align}
\frac{1}{M}\sum_{\tau=0}^{M-1}\mathbb{E}\{U(\tau)\}\leq & \frac{B+V(G_{max}-H_{min})}{\epsilon}\notag\label{eqn:dpthm-Ubnd1}\\
+ & \frac{\mathbb{E}\{L(U(0))\}}{M\epsilon}\end{align}

Taking the limits of \eqref{eqn:dpthm-Ubnd1} as $M\to\infty$ and
setting $\epsilon=r_{max}$ yields \eqref{eqn:dpthm_ubnd}. Setting
$\epsilon=r_{max}$ is done to minimize the bound, as a particular
choice for $\epsilon$ will only affect the bound calculation and
will not affect the policies in any way. 

To prove the utility bounds, note that the concavity of the frame
quality function $g(f(t))$ and the convexity of the playout distortion
function $h(p(t))$ together with Jensen's inequality implies the
following:

\begin{equation}
\frac{1}{M}\sum_{\tau=0}^{M-1}\mathbb{E}\{g(f(\tau))\}\leq g\bigg(\frac{1}{M}\sum_{\tau=0}^{M-1}\mathbb{E}\{f(\tau)\}\bigg)\label{eqn:dpthm-jensen1}\end{equation}

\begin{equation}
\frac{1}{M}\sum_{\tau=0}^{M-1}\mathbb{E}\{h(p(\tau))\}\geq h\bigg(\frac{1}{M}\sum_{\tau=0}^{M-1}\mathbb{E}\{p(\tau)\}\bigg)\label{eqn:dpthm-jensen2}\end{equation}

If we divide \eqref{eqn:dpthm-bnd1} by $MV$ and rearrange it, we
obtain:

\begin{align}
\frac{1}{M}\sum_{\tau=0}^{M-1}\mathbb{E}\{g(f(\tau))\}-\frac{1}{M}\sum_{\tau=0}^{M-1}\mathbb{E}\{ & h(p(\tau))\}\notag\label{eqn:dpthm-ghbnd}\\
\geq(g(f_{\epsilon}^{*})-h^{*})-\frac{B}{V}- & \frac{\mathbb{E}\{L(U(0))\}}{MV}\end{align}

To obtain the frame quality bound, we use the fact that $h(p(t))\geq H_{min}$
in \eqref{eqn:dpthm-ghbnd} and rearrange to get:

\begin{align}
\frac{1}{M}\sum_{\tau=0}^{M-1}\mathbb{E}\{g(f(\tau))\} & \notag\label{eqn:dpthm-fbnd1}\\
\geq(g(f_{\epsilon}^{*})-h^{*})+ & H_{min}-\frac{B}{V}-\frac{\mathbb{E}\{L(U(0))\}}{MV}\end{align}

Using \eqref{eqn:dpthm-jensen1} and taking the limits of \eqref{eqn:dpthm-fbnd1}
as $M\to\infty$ yields:

\begin{equation}
\liminf_{M\to\infty}g(\bar{f})\geq(g(f_{\epsilon}^{*})-h^{*})+H_{min}-\frac{B}{V}\end{equation}

Again, the above bound can be maximized by taking the limit as $\epsilon\to0$.
This produces the frame quality bound \eqref{eqn:dpthm_gbnd}.

To obtain the playout distortion bound, we use the fact that $g(x(t))\leq G_{max}$
in \eqref{eqn:dpthm-ghbnd} and rearrange to get:

\begin{align}
\frac{1}{M}\sum_{\tau=0}^{M-1}\mathbb{E}\{ & h(p(\tau))\}\label{eqn:dpthm-pbnd1}\\
\leq\frac{B}{V}-(g(f_{\epsilon}^{*}) & -h^{*})-G_{max}+\frac{\mathbb{E}\{L(U(0))\}}{MV}\end{align}

Using \eqref{eqn:dpthm-jensen2} and taking the limits of \eqref{eqn:dpthm-pbnd1}
as $M\to\infty$ yields:

\begin{equation}
\limsup_{M\to\infty}h(\bar{p})\leq\frac{B}{V}-(g(f_{\epsilon}^{*})-h^{*})-G_{max}\end{equation}

Taking the limit as $\epsilon\to0$ maximizes the above bound and
produces the frame quality bound \eqref{eqn:dpthm_hbnd}.
\end{IEEEproof}

\subsection{Delay Constrained Optimization Stability and Performance Bounds\label{sub:Delay-Constrained-bnd}}

To obtain the performance bound for the delay constrained Lyapunov
optimization policies in \eqref{eqn:Xlyapobj}, we develop a policy
to enforce a limit on the maximum operating size of the delay accumulator.
To determine such a policy, we observe that for the decoder objective
in \eqref{eqn:Xlyapobj} (last term) to be non-negative, we need:

\begin{align}
U(t)p(t)-Vh(p(t))-X(t)p(t)\geq0\\
\Longrightarrow U(t)p(t)-X(t)p(t)\geq0\\
\Longrightarrow X(t)\leq U(t)\end{align}

Therefore, we introduce the following policy:

\begin{align}
\textrm{If}\quad & X(t)\leq\underbar{U}\notag\nonumber \\
 & \textrm{solve }p(t)\textrm{ by maximizing the last term of \eqref{eqn:Xlyapobj}}\notag\\
\textrm{else} & \notag\label{eqn:del_admctrl}\\
 & p(t)=p_{n}+t_{d}\end{align}

Where $\underbar{U}$ is a positive constant that represents how easily
the maximum operating size of $X(t)$ gets enforced, $\underbar{U}$
is set to 100 in our experiments. \eqref{eqn:del_admctrl} together
with \eqref{eqn:Xbuff} ensures that $X(t)$ does not accumulate anymore
in subsequent timeslots. With \eqref{eqn:del_admctrl}, we introduce
the following corollary:
\begin{cor}
\label{cor:delcon-perfbound}If the network delay variation $r(t)$
is i.i.d over the timeslots, the frame quality function is bounded
as $g(f(t))\leq G_{max}$ and the playout distortion function is bounded
as $h(p(t))\geq H_{min}$. Then implementing the optimization policies
from \eqref{eqn:Xlyapobj} in each timeslot stabilizes the discontinuity
penalty $U(t)$ and satisfies the following performance bounds:

\begin{equation}
\limsup_{M\to\infty}\frac{1}{M}\sum_{\tau=0}^{M-1}\mathbb{E}\{U(\tau)\}\leq\frac{B+\widetilde{C}+V(G_{max}-H_{min})}{p_{max}}\label{eqn:delcor_ubnd}\end{equation}

\begin{equation}
\liminf_{M\to\infty}g(\bar{f})\geq(g^{*}-h^{*})+H_{min}-\frac{B+\widetilde{C}}{V}\label{eqn:delcor_gbnd}\end{equation}

\begin{equation}
\limsup_{M\to\infty}h(\bar{p})\leq\frac{B+\widetilde{C}}{V}-(g^{*}-h^{*})-G_{max}\label{eqn:delcor_hbnd}\end{equation}

Where $V>0$, $g^{*}$, $r^{*}$, $\bar{f}$ and $\bar{p}$ are defined
as in theorem \ref{thm:discontinuity-penalty-policies}. B is defined
as in \eqref{eqn:driftB} and $\tilde{C}$ is defined using $C$ as:

\begin{equation}
\tilde{C}=C+p_{max}(\underbar{{U}}+p_{max})\label{eqn:delcor_ctilde}\end{equation}

Furthermore, implementing limit enforcing policy \eqref{eqn:del_admctrl}
on the delay accumulator $X(t)$ would result in it deterministically
upper bounded for all timeslots $t$ as:

\begin{equation}
X(t)\leq\underbar{{U}}+p_{max}\label{eqn:delcor_xbnd}\end{equation}
\end{cor}
\begin{IEEEproof}
Bound \eqref{eqn:delcor_xbnd} is proved by induction. It can be easily
seen that \eqref{eqn:delcor_xbnd} is satisfied at time 0. Assume
that \eqref{eqn:delcor_xbnd} holds at the current time $t\geq0$,
then we need to prove that $X(t+1)\leq\underbar{{U}}+p_{max}$ in
the next timeslot $t+1$. We have two cases:

1. $X(t)\leq\underbar{{U}}$ : Here $X(t+1)\leq\underbar{{U}}+p_{max}$.
Since from \eqref{eqn:Xbuff}, the maximum delay added to $X(t)$
in one timeslot is $p_{max}$.

2. $X(t)>\underbar{{U}}$ : In this case, the limiting policy \eqref{eqn:del_admctrl}
will be triggered and $X(t)$ will not increase in time $t+1$ resulting
in $X(t+1)\leq X(t)\leq\underbar{{U}}+p_{max}$.

This proves \eqref{eqn:delcor_xbnd}. To prove \eqref{eqn:delcor_ubnd},
\eqref{eqn:delcor_gbnd} and \eqref{eqn:delcor_hbnd}, observe that
using \eqref{eqn:delcor_xbnd}:

\begin{align}
X(t) & \mathbb{E}\{p_{n}+t_{d}-p(t)|U(t),X(t)\}\notag\\
\geq & -X(t)\mathbb{E}\{p(t)|U(t),X(t)\}\notag\\
\geq & -(\underbar{{U}}+p_{max})p_{max}\label{eqn:delcor_minibnd}\end{align}

Using \eqref{eqn:delcor_minibnd} and \eqref{eqn:delcor_ctilde} in
\eqref{eqn:Xlyapobj} will yield:

\begin{align}
\Delta(U,X) & -V\mathbb{E}\{g(f)-h(p)|U,X\} & \notag\label{eqn:delcor_Xlyapobj}\\
\leq B & +\tilde{{C}}\notag\\
 & -\mathbb{E}\{Vg(f)-UF(f)|U,X\} & \notag\\
 & -\mathbb{E}\{Up-Vh(p)-Xp|U,X\}\end{align}

The proof then proceeds exactly as in theorem \ref{thm:discontinuity-penalty-policies}.
\end{IEEEproof}

\subsection{Stability and Performance Bounds of Policies with Network Delays
\label{sub:Perf-Bnds-del}}
\begin{cor}
\label{cor:apx-delstab}If the network delay variation $r(t)$ is
i.i.d over the timeslots, the frame quality function is bounded as
$g(f(t))\leq G_{max}$ and the playout distortion function is bounded
as $h(p(t))\geq H_{min}$. Then implementing the optimization policies
from \eqref{eqn:delcon-deriv2} in each timeslot stabilizes the discontinuity
penalty $U(t)$ and satisfies the following performance bounds:

\begin{equation}
\limsup_{M\to\infty}\frac{1}{M}\sum_{\tau=0}^{M-1}\mathbb{E}\{U(\tau)\}\leq\frac{B'''+V(G_{max}-H_{min})}{p_{max}}\end{equation}

\begin{equation}
\liminf_{M\to\infty}g(\bar{f})\geq(g^{*}-h^{*})+H_{min}-\frac{B'''}{V}\end{equation}

\begin{equation}
\limsup_{M\to\infty}h(\bar{p})\leq\frac{B'''}{V}-(g^{*}-h^{*})-G_{max}\end{equation}

Where $V>0$, $g^{*}$, $r^{*}$, $\bar{f}$ and $\bar{p}$ are defined
as in theorem \ref{thm:discontinuity-penalty-policies}. $B'''$ is
defined as:

\begin{equation}
B'''=B'+d_{b}(d_{f}+1)\left(r_{max}^{2}+r_{max}p_{max}\right)\label{eq:apx-cor-delstab-B}\end{equation}

with $B'$ from \eqref{eqn:driftBdelay}.\end{cor}
\begin{IEEEproof}
We first look at \eqref{eqn:drift4}. Note that with the recursive
definition of $U(t)$ \eqref{eqn:ut}, \eqref{eqn:drift4} can be
rewritten as:

\begin{align}
\Delta(U) & \leq\; B''-U\mathbb{E}\left\{ \gamma_{d_{f}}\big|U\right\} -\mathbb{E}\left\{ -U_{d_{b}}F_{d_{f}}\big|U\right\} \notag\\
 & \leq\; B''-\mathbb{E}\left\{ \left(\gamma_{d_{f}}\left[U_{d_{b}}-\gamma\right]^{+}+F\gamma_{d_{f}}\right)\big|U\right\} -\mathbb{E}\left\{ -U_{d_{b}}F_{d_{f}}\big|U\right\} \end{align}

Since $\left[U_{d_{b}}-\gamma\right]^{+}\leq U_{d_{b}}$, the above
equation becomes:

\begin{equation}
\Delta(U)\leq\; B''-\mathbb{E}\left\{ \left(\gamma_{d_{f}}U_{d_{b}}+F\gamma_{d_{f}}\right)\big|U\right\} -\mathbb{E}\left\{ -U_{d_{b}}F_{d_{f}}\big|U\right\} \label{eq:apx-cor-delstab1}\end{equation}

Note that the term $F\gamma_{d_{f}}$ can be upper bounded as:

\begin{equation}
F\gamma_{d_{f}}\leq d_{b}(d_{f}+1)r_{max}p_{max}\label{eq:apx-cor-delstab2}\end{equation}

Using \eqref{eq:apx-cor-delstab2} in \eqref{eq:apx-cor-delstab1}
yields:

\begin{equation}
\Delta(U)\leq\; B'''-\mathbb{E}\left\{ \gamma_{d_{f}}U_{d_{b}}-U_{d_{b}}F_{d_{f}}\big|U\right\} \end{equation}

with $B'''$ defined in \eqref{eq:apx-cor-delstab-B}. This then implies
that \eqref{eqn:delcon-deriv1} can be bounded as:

\begin{align}
\notag\Delta(U) & -V\mathbb{E}\{g(f(t))-h(p(t))|U\}\\
\notag & \leq\; B''-\mathbb{E}\left\{ U\gamma_{d_{f}}-U_{d_{b}}F_{d_{f}}\big|U\right\} \\
\notag & \quad\quad\quad-\mathbb{E}\left\{ Vg(f(t))-Vh(p(t))\big|U\right\} \\
\notag & \leq\; B'''-\mathbb{E}\left\{ U_{d_{b}}\gamma_{d_{f}}-U_{d_{b}}F_{d_{f}}\big|U\right\} \\
 & \quad\quad\quad-\mathbb{E}\left\{ Vg(f(t))-Vh(p(t))\big|U\right\} \end{align}

The proof then proceeds as in theorem \ref{thm:discontinuity-penalty-policies}.
\end{IEEEproof}

\end{document}